\documentclass[10pt,letterpaper]{article}
\usepackage[top=0.85in,left=2.75in,footskip=0.75in]{geometry}

\usepackage{amsmath,amssymb}

\usepackage{changepage}

\usepackage{textcomp,marvosym}

\usepackage{cite}

\usepackage{nameref,hyperref}

\usepackage[nopatch=eqnum]{microtype}
\DisableLigatures[f]{encoding = *, family = * }

\usepackage[table]{xcolor}

\usepackage{array}

\newcolumntype{+}{!{\vrule width 2pt}}

\newlength\savedwidth

\raggedright
\usepackage[aboveskip=1pt,labelfont=bf,labelsep=period,justification=raggedright,singlelinecheck=off]{caption}

\makeatletter
\renewcommand{\@biblabel}[1]{\quad#1.}
\makeatother

\usepackage{lastpage,fancyhdr,graphicx}
\usepackage{epstopdf}
\fancyheadoffset[L]{2.25in}
\fancyfootoffset[L]{2.25in}
\begin{document}
\vspace*{0.2in}

\begin{flushleft}
{\Large
\textbf\newline{User authentication system based on human exhaled breath physics} 
}
\newline
\\
Mukesh Karunanethy\textsuperscript{1},
Rahul Tripathi\textsuperscript{1},
Mahesh V Panchagnula\textsuperscript{1*},
Raghunathan Rengaswamy\textsuperscript{2}
\\
\bigskip
\textbf{1} Department of Applied Mechanics and Biomedical Engineering, Indian Institute of Technology Madras, Chennai, Tamil Nadu, India
\\
\textbf{2} Department of Chemical Engineering, Indian Institute of Technology Madras, Chennai, Tamil Nadu, India
\\
\bigskip

%
%





* mvp@iitm.ac.in

\end{flushleft}
\section*{Abstract}
This work, in a pioneering approach, attempts to build a biometric system that works purely based on the fluid mechanics governing exhaled breath. We test the hypothesis that the structure of turbulence in exhaled human breath can be exploited to build biometric algorithms. This work relies on the idea that the extrathoracic airway is unique for every individual, making the exhaled breath a biomarker. Methods including classical multi-dimensional hypothesis testing approach and machine learning models are employed in building user authentication algorithms, namely user confirmation and user identification. A user confirmation algorithm tries to verify whether a user is the person they claim to be. A user identification algorithm tries to identify a user's identity with no prior information available. A dataset of exhaled breath time series samples from 94 human subjects was used to evaluate the performance of these algorithms. The user confirmation algorithms performed exceedingly well for the given dataset with over $97\%$ true confirmation rate. The machine learning based algorithm achieved a good true confirmation rate, reiterating our understanding of why machine learning based algorithms typically outperform classical hypothesis test based algorithms. The user identification algorithm performs reasonably well with the provided dataset with over $50\%$ of the users identified as being within two possible suspects. We show surprisingly unique turbulent signatures in the exhaled breath that have not been discovered before. In addition to discussions on a novel biometric system, we make arguments to utilise this idea as a tool to gain insights into the morphometric variation of extrathoracic airway across individuals. Such tools are expected to have future potential in the area of personalised medicines.



\section*{Introduction}
Human exhaled breath is largely turbulent. During exhalation, air is forced out of the lung through trachea by the contracting diaphragm. To start with, the Reynolds number associated with flow through trachea is sufficiently high. In addition, as the air passes through the trachea, it interacts with the complex internal structures associated with the upper respiratory tract, leading to complexity in the flow. The upper respiratory tract consists of the larynx, the pharynx, and the oral cavity. Owing to the complexity associated with the interaction between air that is already turbulent with the upper respiratory tract, we hypothesize that the turbulent signatures in the exhaled air are unique and identifiable from person-to-person. A plausible way to test this hypothesis is to build a user authentication system that would answer the question of classifiability of a human subject purely based on the fluid dynamics of the exhaled breath, essentially serving the purpose of a biometric user authentication system. Such a system is a real-time system to verify a user's identity using any measured feature pertaining to the user's physiology or behaviour. Thus, authentication can be broadly seen as comprising two classes of methods: \textit{physiological biometrics} (eg., fingerprints, iris scans, facial recognition, etc.) and \textit{behavioural biometrics} (eg., gait analysis, voice ID, breathing gesture (\cite{chauhan2017breathprint}), etc.). There are two major modes of deployment of a user authentication/access system are: $(i)$ \textit{user confirmation}, and $(ii)$ \textit{user identification}. In the confirmation mode, a user declares his or her identity, which is to be confirmed. In this case, the user's biometric data is compared to a specific set of data of the same person obtained during an enrollment process. In the identification mode, a user does not disclose his or her identity. In that case, a user's data is compared with all registered data in the database of bona fide users, and the user is identified. We will discuss algorithms for testing the two biometric modes in this manuscript and argue that exhaled breath contains sufficient information to implement both biometric modes.

Human exhaled breath has proven to be a non-invasive diagnostic tool for a spectrum of medical problems as well. \cite{schaber2018breathprinting} studied the diagnosis of malarial infection by analysing the breath composition, or ``breathprint'' which contains a series of volatile organic compounds (VOCs) produced by the \textit{P. falciparum}-infected erythrocytes. They built a nearest mean binary classifier with leave-1-breath-sample-out cross-validation scheme to assign predictions. The European Respiratory Society (ERS) technical standard (\cite{horvath2017european}) reported that the fraction of nitric oxide in exhaled gas is a potential biomarker for lung diseases. \cite{rattray2014taking} showed the potential of breath-based metabolomics (breathomics) in personalised medicine. Mass spectrometry is one of the main platforms used for data profiling in these techniques. In their study, \cite{samara2013single} reported enhancements required in the analysis of single exhaled breath metabolomic data for the unique identification of patients with acute decompensated heart failure. \cite{guo2010diabetes} made attempts to develop a breath analyzer system to measure blood glucose levels and to classify diabetic/non-diabetic patients using a support vector machine (SVM) classifier based on acetone levels in breath measured using chemical sensors. \cite{lawal2017exhaled} reviewed various breath sampling methods with a bibliometric study. \cite{mashir2009exhaled}, \cite{pereira2015breath}, and \cite{das2020non} studied the potential advantages of breath tests as a non-invasive technique with potential biomarkers in disease diagnosis. The above efforts in the literature proving exhaled breath as a biomarker largely involve the analysis of its chemical composition by various techniques. In other words, these studies have shown that the compounds present in exhaled air produce a molecular signature. There exists no evidence in the literature of any attempt to develop an identifier purely based on the fluid dynamic aspects of the exhaled airflow.

Respiratory flow measurements are widely performed using spirometers and pneumotachographs. Inspirational flow patterns in humans were studied using measurements from a cycloergometer to theoretically estimate mechanical work during inhalation by \cite{lafortuna1984inspiratory}. \cite{painter1992analyses} studied the human respiratory flow patterns using pneumotachographic flow measurements at the mouth. Hot wire anemometry (HWA) has been used by several researchers in the past for respiratory flow measurements. \cite{godal1976application} demonstrated the application of HWA in respiratory flow measurements in small animals. \cite{Lundsgaard1979tc} investigated the performance of a constant temperature hot wire anemometer (CT-HWA) system for respiratory gas flow rate measurements. The study demonstrated that a CT-HWA will meet the response requirements and be insensitive to changes in temperature and humidity that are frequently experienced in respiratory flows. In the research by \cite{silva2002} and later by \cite{araujo2004breathing}, it was shown that CT-HWA can be used to measure fluid flow in the forced oscillations technique applied to the human respiratory system, as a substitute for the pneumotachograph.  Other studies reporting the implementation of CT-HWA for measuring expiratory flow parameters are by \cite{kandaswamy2002ieee} and \cite{xu2015indoor}. \cite{plakk1998hot} showed that CT-HWA can be used as a flow transducer for spirography. In conclusion, HWA is a robust tool for obtaining time-resolved turbulence signature measurements in flows. Most of the work in the literature has taken advantage of the HWA data for flow rate calculations, effectively using it only as an alternative for spirometry-based studies. We propose to use HWA measurements (the complete time series of instantaneous velocity data) of turbulence in human exhaled breath as input signals for the development of a biometric system.

Behavioural biometrics use a person's gestures, such as gait patterns or breathing gestures. Recent work by \cite{chauhan2017breathprint,chauhan2018breathing} revealed the prospects of exploiting breathing acoustics for user authentication. They built a new behavioural biometric signature called \textit{BreathPrint} based on audio features acquired from a microphone sensor in smartphones, wearables and other IoT devices. \cite{chauhan2017breathprint} deployed a conventional machine learning model based on the Gaussian mixture model (GMM), while \cite{chauhan2018breathing} established the feasibility and performance evaluation of RNN-based deep learning models. A novel WiFi-based breathing estimator \textit{UbiBreathe} developed by \cite{abdelnasser2015ubibreathe} works as a respiratory monitoring system based on the received signal strength (RSS) data from a nearby WiFi-enabled device. A continuous user verification system was developed using this approach by \cite{liu2020continuous} for round-the-clock user verification, built based on user-specific respiratory features derived based on waveform morphology analysis and fuzzy wavelet transformation. A deep learning-based scheme also detects the existence of spoofing attacks. \cite{lu2020} developed a speaker recognition system, \textit{BreathID} based on breath biometrics. Breath during speech is considered trivial or a noise component. They showed that unique breath features can be formulated by a template matching technique for speaker recognition.

In summary, the use of HWA and, more broadly, \textit{breath turbulence measurements as a tool for biometric authentication} has not been attempted in the literature. Conventional biometric systems such as voice, face, and fingerprint recognition have their own disadvantages. There is a need to develop more sophisticated biometric systems that could make use of internal physiological features of the human body. We attempt to build a novel user authentication system based on human exhaled breath, using the principles of multidimensional hypothesis testing and machine learning. This system is different from an acoustics-based biometric system, since it does not require vocal data from the human subject and is built solely on the fluid dynamic information contained in the exhaled breath.

\section*{The experimental dataset and methodology}
\begin{figure}[!h]
\includegraphics[scale=0.45]{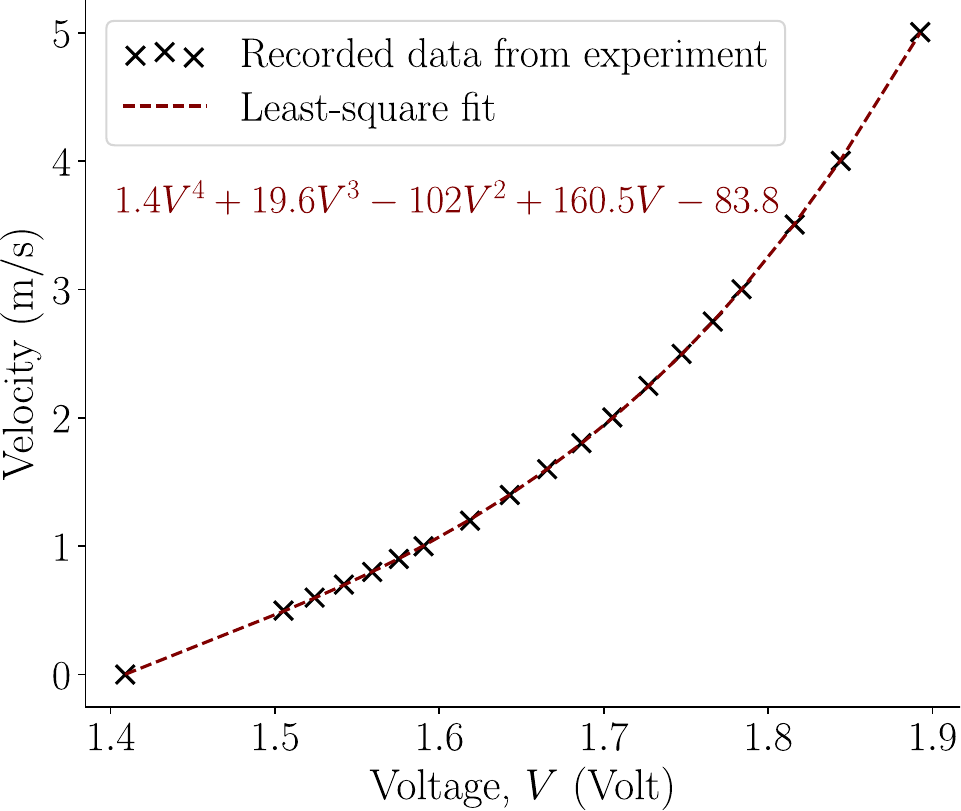}
\caption{{\bf Calibration curve for the hot wire anemometer.}
A fourth order least square fit of the experimental data (shown as maroon dotted line) becomes the \textit{calibration curve} for the hot wire anemometer in use. The polynomial equation of the fourth order fit is shown inside the plot.}
\label{fig:calibration_fit}
\end{figure}

A measurement-based study was employed to develop algorithms for biometric authentication. Measurements of the exhaled breath were made using a Dantec Dynamics\textsuperscript{\tiny\textregistered} $55\mathrm{P}11$ hot wire probe. It consists of a $5\mu\mathrm{m}$ diameter, $1.25\mathrm{mm}$ long platinum-coated tungsten wire, which acts as the sensor. A Dantec Dynamics MiniCTA\textsuperscript{\tiny\textregistered} $54\mathrm{T}42$ module housed the CT-HWA's signal processing and output system. The hot wire probe was calibrated using a standard procedure of simultaneous measurement of the flow velocity and the anemometer voltage. The calibration was performed using a Dantec Dynamics StreamLine Pro\textsuperscript{\tiny\textregistered} automatic calibrator, between a velocity range of $0-5~\mathrm{m/s}$. Using this procedure, we were able to determine the calibration constants from an assumed velocity-voltage relation. This relation is a least-square polynomial fit of order-$4$ in the velocity-voltage space as shown in Fig \ref{fig:calibration_fit}. In the current study, the raw voltage time series was itself used in all the analysis. This helps us avoid frequent re-calibration of the probe. The initial calibration was performed only to make sure that the voltage and velocity signals are monotonically positively correlated (as can be inferred from the least square fit from Fig \ref{fig:calibration_fit}).

\subsection*{Participants}
94 participants were recruited to take part in this study, following the ethical approval from the Institutional Ethics Committee (IEC) of the Indian Institute of Technology Madras, Chennai, India (IITM - IEC Protocol No. IEC/2018-03/MP/01). The participants were students of the Indian Institute of Technology Madras. Their age ranged from 21 to 27 years. Data were collected only once (one set of 10 breath samples) per participant. Volunteers with epileptic disorder were excluded from participation. The experimental data collection was carried out between 8th and 17th January, 2019. All volunteers who participated in this study had given their written consent. The recorded time series data were analyzed anonymously.

\subsection*{Data collection and analysis}
\begin{figure}[!h]
\includegraphics[scale=0.136]{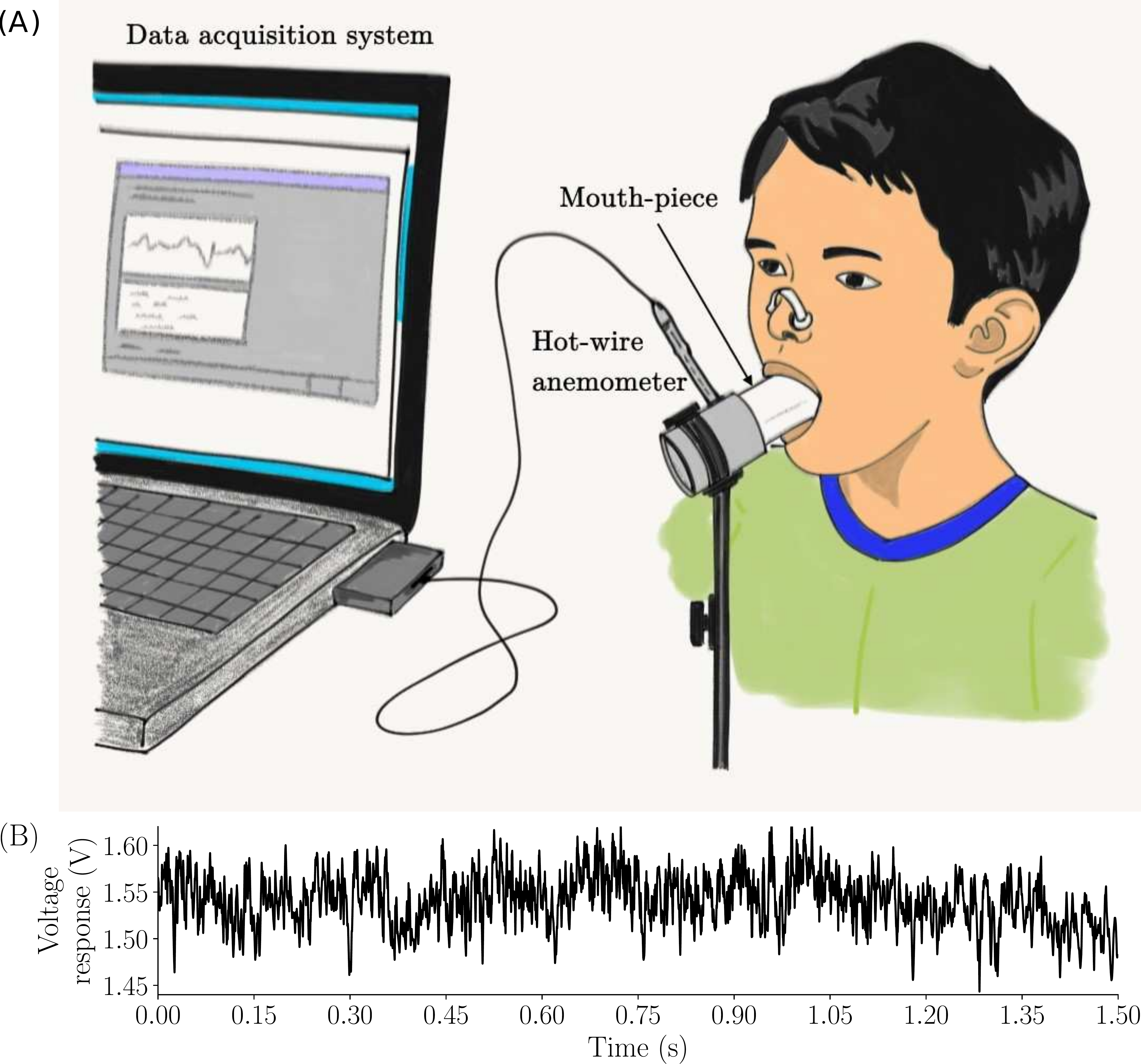}
\caption{{\bf Experimental setup and recorded time series.}
(A) Depiction of the experimental setup for data collection. It consists of a disposable mouth-piece, a mouth-piece mount housing a hot wire anemometer and a data acquisition system. (B) A typical human exhalation velocity signal measured using a standard hot wire anemometer. The time signals were sampled at $10\mathrm{kHz}$ for $1.5~\mathrm{seconds}$.}
\label{fig:data_acquisition}
\end{figure}

A schematic of the experimental setup is shown in Fig $\ref{fig:data_acquisition}$A. It consists of a mouthpiece assembled into an aluminium circular cross-section channel which housed the hot-wire probe aligned to its axis to measure the streamwise component of the turbulent exhaled flow velocity.  The human subjects were allowed to exhale through their mouths into the experimental measurement setup. The nose was clipped during data recording to ensure that all the exhaled air passes through the oral cavity before entering the experimental setup. Each human subject was provided with a new disposable plastic mouth-piece to wrap their mouth around, through which the subjects exhaled. The obstruction of the tongue to the flow was avoided by placing the mouth-piece above the tongue. Data were obtained in each exhalation trial lasting about 1.5 seconds, with 10 trials recorded per subject. Each time series was recorded by sampling the voltage response at $10\mathrm{kHz}$. This effectively gives us $15000$ data points in a time series, the relevance of which would be discussed in the following sections. The time series signal from a typical exhalation trial is shown in Fig $\ref{fig:data_acquisition}$B. Given a set of time series signals from a library of users, our algorithm comprises of segmentation, normalization, feature extraction and subdivision of feature set into training and testing sets. The training dataset became part of the enrolled database, whereas the testing dataset was used for testing the performance of the authentication algorithms. The enrollment and algorithm testing depends on the type of algorithm being used. More details of user authentication systems are discussed in section titled \textit{User confirmation algorithms}.

\subsection*{Statistical description of the time series}
In general, a statistical description would involve the representation of time series distributions in terms of the central moments. Such representative measures tend to vary within a non-stationary time series. They can be characterized by studying how these moments depend on time intervals within the time series itself, by investigating the scaling properties of the signal. For instance, the Hurst exponent, $\mathrm{H}$ (\cite{Hurst1951}) parametrizes the effect of the statistics of time intervals on the standard deviation of the time signal. In the context of multifractal analysis, the generalized Hurst exponent $\mathrm{H}(q)$ is used for parameterization, where $q$ is the order of the fluctuation function (\cite{kantelhardt2002multifractal}). $\mathrm{H}(q)$ is also known as the $q-$order Hurst exponent. In our study, we focus on the multifractal properties of the time series, since interestingly, human exhaled breath has been found to display multifractality, based on our analysis which will be discussed in this section. Fully developed turbulence is known to exhibit multifractality, as described by \cite{Sreenivasan1991}.

The multifractal nature of exhaled breath signals were investigated using the well-known technique called multifractal detrended fluctuation analysis (MFDFA) developed by \cite{kantelhardt2002multifractal}. It helps us to identify multifractal scaling properties as well as to detect long-range correlations in a time series. A detailed explanation of the theory behind this algorithm can be found in the original work by \cite{kantelhardt2002multifractal}. A step-by-step implementation of the MFDFA program using Matlab\textsuperscript{\tiny\textregistered} was given by \cite{Ihlen2012}. We made use of the recommendations from the \cite{kantelhardt2002multifractal} and \cite{Ihlen2012} to write a \textit{Python} program to perform the MFDFA on exhaled breath time signals. Briefly, the algorithm involves dividing the time series data into time intervals of equal length, then applying detrended fluctuation analysis (DFA) (\cite{Peng1994}) to each time interval to remove the trend and then calculating the fluctuation function $\mathrm{F}$. Next, the $q-$order fluctuation function $\mathrm{F}(q)$ is obtained by raising the detrended fluctuation function to the power of $q$. The $q-$order Hurst exponent $\mathrm{H}(q)$ is obtained from the scaling behavior of $\mathrm{F}(q)$. Then, the algorithm involves estimating the $q-$order mass exponents $\tau (q)$ from $q-$order Hurst exponent $\mathrm{H}(q)$, converting them into the $q-$order singularity exponents $\alpha$, and then computing the generalized singularity dimensions, also known as the singularity spectrum $f(\alpha)$.

In the context of multifractal analysis, a measure of complexity of a time series is its singularity spectrum $f(\alpha)$, which characterizes the distribution of fractal dimensions or scaling exponents $\alpha$ across different parts of the signal. While conventional DFA (\cite{Peng1994}) quantifies the average correlation properties of a signal purely through the scaling exponent $\alpha$, MFDFA provides another important measure.  The width of the multifractal spectrum $\omega$ (see Fig \ref{fig:multifractal_spectrum}), which indicates the richness of multifractality present in the experimental data adds further insight into the data. Third-order polynomial fits were used to detrend data in each time interval. The time interval (window) sizes range between $10$ and $N/4$ data points, where $N$ is the length of the time series. The orders $q$ of fluctuation function ranges from $-5$ to $5$. It is to be noted that the input time series to the analysis was first normalized, which is discussed in a subsequent section. The chosen normalization method does not alter the compact support of the input time series, as it is essential that a time series with compact support is necessary for reliable multifractal analysis.

\begin{figure}[!h]
\includegraphics[scale=0.175]{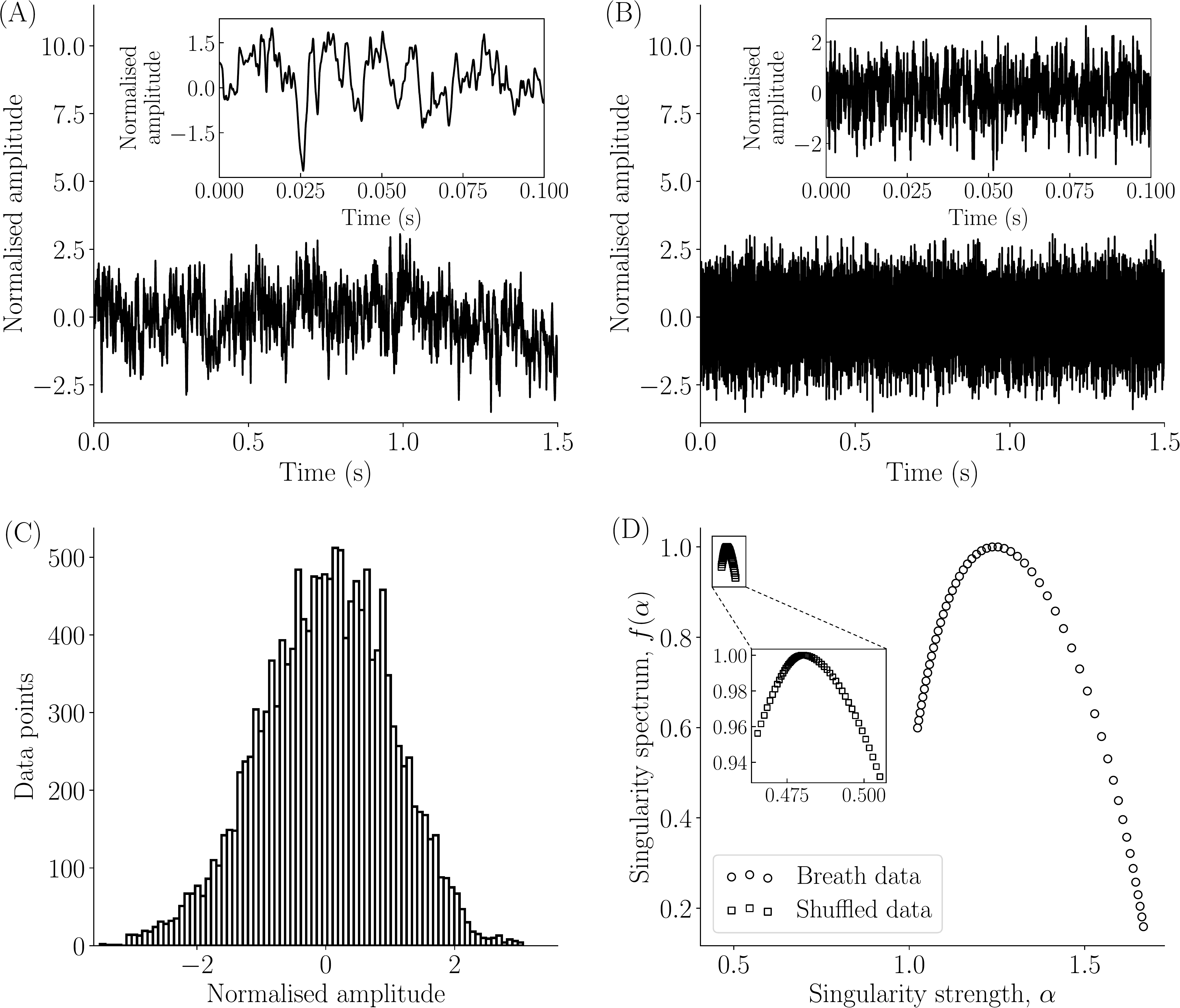}
\caption{{\bf Comparison of multifractality of time signals.}
Plots showing the effect of random shuffling of exhaled breath time series acquired using a hot wire anemometer. Signals shown in (A) and (B) correspond to the actual breath data and the shuffled data respectively. Inset plots in (A) and (B) show a zoomed-in view of the first $1000$ data points of the signals. Note that the signal has been normalized using its mean and standard deviation. (C) Histogram showing the distribution of all $N$ data points of the breath signal. (D) Multifractal spectra for the original breath signal and the randomly shuffled white noise signal. Random shuffling causes loss of memory within the time series and losses the multifractality.}
\label{fig:multifractal_signal_comparison}
\end{figure}

Fig \ref{fig:multifractal_signal_comparison} consists of a set of plots showing the effect of random shuffling of the exhaled breath time signal on the multifractal singularity spectrum. Figs \ref{fig:multifractal_signal_comparison}A and \ref{fig:multifractal_signal_comparison}B show the original and shuffled time series respectively. The inset plots in each of these plots display the zoomed-in view of the first $1000$ data points. It is clearly visible that the existing correlation is destroyed when the data is shuffled. The distribution of the visualised time signal is shown in the form of a histogram in Fig \ref{fig:multifractal_signal_comparison}C.

The Kolmogorov–Smirnov test for normality (\cite{Massey1951}) revealed that a large fraction of the available breath signals were non-Gaussian. Deviations from a Gaussian or symmetric distribution may be a sign of multifractality stemming from a broad probability distribution function (PDF) as described by \cite{kantelhardt2002multifractal}. Shuffling the time series helps us in discovering the reason for multifractality in this case. By randomly permuting the order of values in the time series, temporal correlations are disrupted while preserving the PDF. If the multifractality persists in the shuffled or surrogate data, it suggests that the broad PDF is the primary source of multifractality. Conversely, if the multifractality disappears in the shuffled data, it indicates multifractality due to inherent long-range temporal correlations. Fig \ref{fig:multifractal_signal_comparison}D is a plot of the singularity spectral function $f(\alpha)$ against the singularity strength $\alpha$, resulting from the MFDFA on the time series from \ref{fig:multifractal_signal_comparison}A and \ref{fig:multifractal_signal_comparison}B. The plot consists of two representative multifractal spectra - one for the exhaled breath time series and the other corresponding to the same time series shuffled, which becomes a white noise. The white noise signal was observed to form only a tiny arc clustered around $\alpha=0.5$, while the multifractal breath signal forms a well-defined spectrum. This observation is evidence of the presence of long-range correlations in the breath time signal. Any memory of the correlations (strong or weak) within the time series is lost when randomly shuffled. The inset plot in Fig \ref{fig:multifractal_signal_comparison}D shows a magnified view of the spectrum from the white noise signal. It can be inferred from this observation that the white noise signal does not show any degree of multifractality and also reconfirms that the multifractality of exhaled velocity is defined by its inherent long-range correlation properties, both for short- and long-range fluctuations. The multifractal analysis was made use in the time series segmentation and feature extraction, which will be discussed in the following sections.

\subsection*{Time series segmentation, normalization and selection} \label{sec:segmentation_normalization}
Segmentation of time series is a standard practice in many data analysis techniques to obtain dividing points on a signal with or without stationarity. In machine learning problems with limited availability of time series samples, segmentation is of vital importance. By performing an efficient segmentation on the basis of certain statistical measures, we can obtain sufficient number of samples to train and test machine learning models. Fig \ref{fig:data_acquisition}B is a plot showing the instantaneous voltage response from the hot wire probe for $1.5$ seconds. It was obtained by sampling at a frequency of $10~\mathrm{kHz}$, giving us a sufficiently resolved long series to perform segmentation without losing any significant information on the flow physics. 

We will now discuss the segmentation process. Each time signal was divided into $19$ overlapping segments using a window size equal to $1/10$th the length of the signal and a sliding width of half the segment size. Remember the machine learning models may tend to overfit the training data when there are large number of overlapping segments. The purpose of using overlapping windows was to capture the end effects of the time series segments during feature extraction. So, the chosen segment width and sliding width are justified as each part of the time signal appears only in two segments. This effectively gives $1500$ data points to each segment making it sufficiently long to reliably extract features using tools discussed this manuscript. As a result, a maximum of $190$ representative time blocks become available for the analysis for each user. Each of the time signals were normalized before feature extraction, making the time series comparable across realizations. This would also make all signals independent of the sensor being used for the measurement, since these features only rely on the temporal correlation structure in the series and not on the raw data values. This approach can be termed as being \textit{sensor-agnostic}. Regardless of whether the time series signal is measured using a hot-wire/film probe, or a laser-based technique, the performance of the algorithm will not be affected, as long as there are sufficient data points to properly capture the temporal structure in the flow. We then build an algorithm which works with these features which are invariant to the absolute value of the time series. $z$-score normalization which is popularly known as standardization was used to normalize the time series. To perform $z$-score normalization, the mean of the entire time series is subtracted from each data point in the time series. Then, the resulting values are divided by the standard deviation of the time series. This scales the data so that it has a mean of zero and a standard deviation of one. The resulting normalized time series will have values that represent the number of standard deviations away from the mean. The $z$-score normalization has the form shown in Eq \ref{eq:normalization}.
\begin{equation} \label{eq:normalization}
    z(i) = \frac{x(i)-\mu_t}{\sigma}, \quad i=1,2,\ldots N
\end{equation}
where $z(i)$ is the normalized time series, $x(i)$ is the original time series of length $N$, $(\mu_t)$ is the mean of the time series, and $(\sigma)$ is the standard deviation of the time series. The time series becomes unitless after normalization.

\begin{figure}[!h]
\includegraphics[scale=0.45]{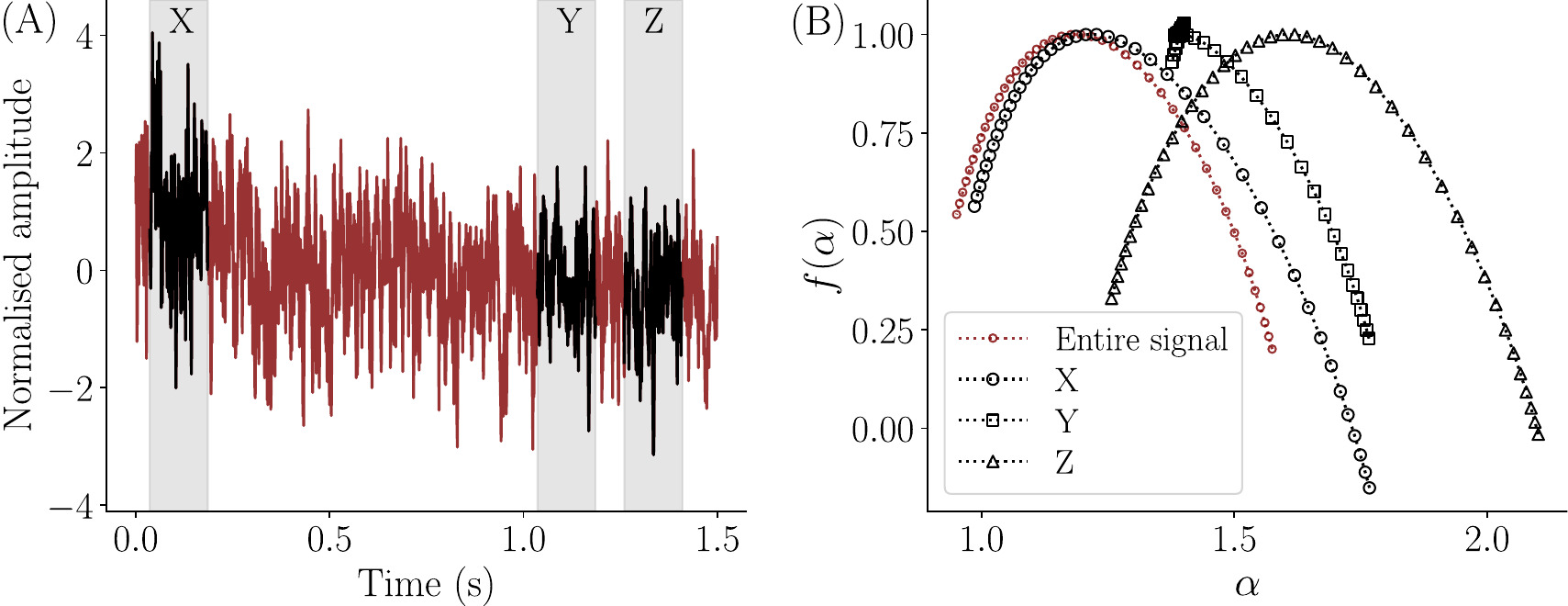}
\caption{{\bf Multifractal spectra for different segments of a time signal.}
The multifractral spectra corresponding to the entire time signal (maroon) and time segments X, Y and Z (black, bounded by gray band) in (A) are shown in (B). It is evident that few segments exhibit an inverted parabola shape and spectrum B has a distortion.}
\label{fig:segment_spectrum}
\end{figure}

MFDFA was performed on all normalised time series, and it revealed that not all spectra exhibit the expected shape. The general shape of a multifractal spectrum is convex or more precisely an inverted parabola, with the peak occurring at the central moment. This convex shape signifies the presence of multifractal scaling, indicating that different parts of the time series exhibit distinct scaling behaviors. Certain time segments were observed to result in a spectrum with folds or distortions. Fig \ref{fig:segment_spectrum} shows an example of such a distortion. The multifractal spectrum for a time signal and three randomly chosen segments X, Y and Z from the same time series are displayed. Fig \ref{fig:segment_spectrum}A shows the entire time signal and the chosen segments. Out of the three segments, X and Z show a typical spectral shape, whereas segment B consists of a fold towards the left hand side of the spectrum (see Fig \ref{fig:segment_spectrum}B).

There could be several reasons for the appearance of folds in the multifractal spectrum. $(i)$ They could occur due to irregularities or data artifacts in the time series itself, such as noises, outliers, etc. which may arise due to inconsistent exhalation by the user during data acquisition. For example, during the period of $1.5$ seconds, if the user exhales abruptly for the first $1$ second of the trial, and then the breath velocity steadily decays for the remaining $0.5$ seconds. The segment which falls between these two regions might contain irregularities within it. Such irregularities could introduce inconsistencies in the scaling behaviour. $(ii)$ The spectrum may be affected by the non-stationarity of the time series, which is when the statistical properties change with time, such as due to change in breath velocity. $(iii)$ Spectral folds might even arise due to the finite size of the time segment. Limited number of data points may not capture the scaling properties at different scales. Investigating the type of distortions or the reason behind this behaviour of the spectrum for certain time segments fell outside the scope of this work. Instead, we made use of this behaviour as an indicator to judge whether a segment is valid or not. All segments which showed non-convex singularity spectra were discarded in our analysis. Also, the segments which produce a spectral width less than $0.05$ were rejected, since they exhibit a very low degree of multifractality. These two strategies effectively make MFDFA a tool for time series selection, for further feature extraction and analysis. Any time signal which contains significant number of segments with inconsistent scaling behaviour can be rejected using this tool during the data recording step itself. A numerical example discussing how a multifractal singularity spectrum can have non-convex shapes can be found in \cite{Eke2012}.

\subsection*{Feature extraction} \label{sec:feature_extraction}
\begin{figure}[!h]
\includegraphics[scale=0.55]{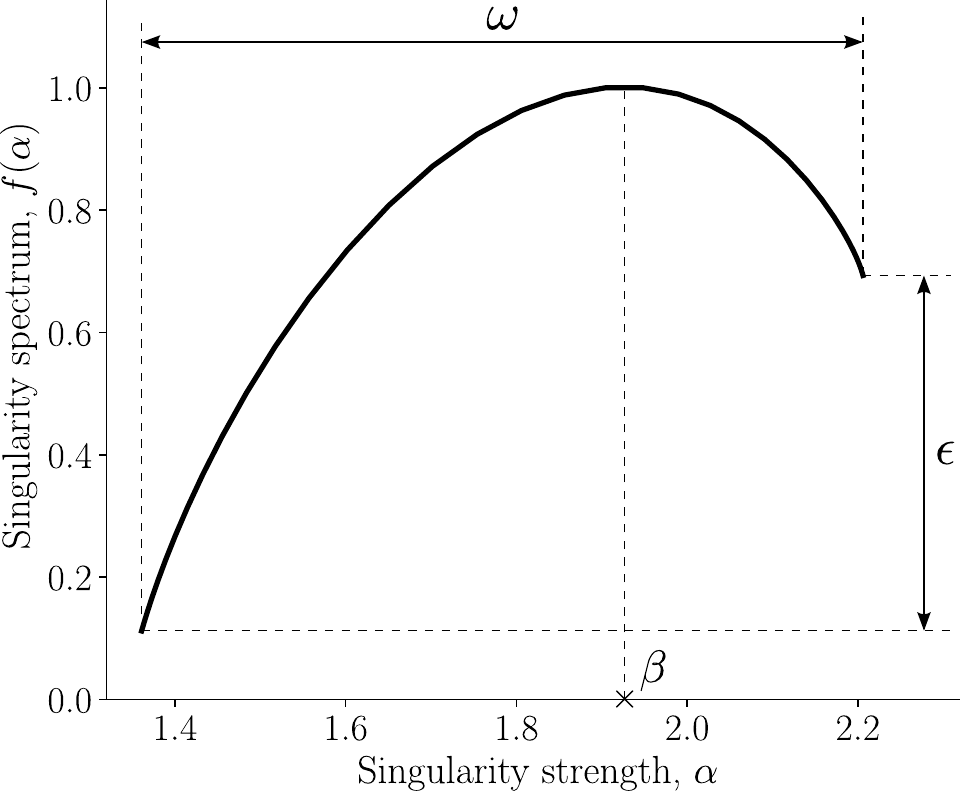}
\caption{{\bf The multifractal spectrum.}
Plot of the spectrum of singularities $f(\alpha)$ against the singularity strength $\alpha$, computed for an exhalation time series segment. The parameters $\beta$, $\omega$ and $\epsilon$ are the features that characterize a multifractal spectrum.}
\label{fig:multifractal_spectrum}
\end{figure}

Features were extracted from normalized time signals using various time series feature extraction techniques. Unlike other physiological biometric systems where image-based patterns or features are used as templates to match an individual's identity, our input data is a time series from an individual which requires feature extraction. Several features of the time series were studied in order to develop insights into the data. The multifractal spectral information was incorporated into our analysis by including them in the set of features. The fact that the time series contains information pertaining to the correlation structure becomes relevant to machine learning algorithms. In keeping with this principle, we extract a set of three important features from the spectrum: $(i)$ $\beta$, the abscissa corresponding to the spectral maxima, $(ii)$ $\omega$, the width of the spectrum, and $(iii)$ $\epsilon$, the bias or asymmetry parameter of the spectrum. The parameters $\beta$, $\omega$ and $\epsilon$ are dimensionless. These features are visualised on the multifractal spectrum of an exhaled breath time signal in Fig \ref{fig:multifractal_spectrum}. It was also noted from our analysis that the spectra showed clear differences in their temporal structure; i.e., parameters such as $\beta$, $\omega$ and $\epsilon$ were different for different time signals. Several other multifractal spectral features have also been considered in the literature (\cite{SHIMIZU2004}, \cite{Eke2012}, \cite{ZHANG2018}). We chose these three features for simplicity and also they encompass most important descriptions of a multifractal spectrum. Investigating how unique these features behave is of interest to this work.

\begin{figure}[!h]
\includegraphics[scale=0.13]{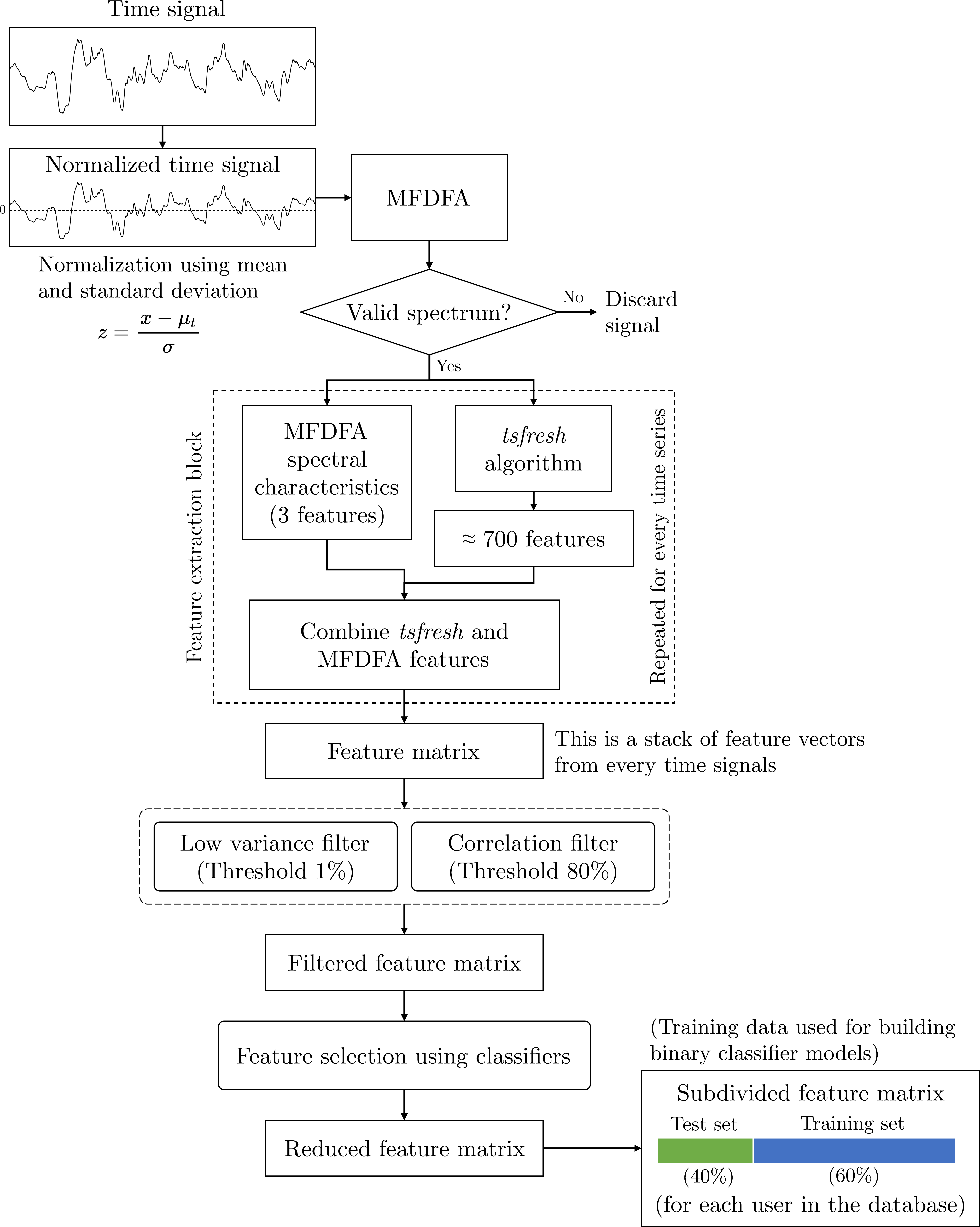}
\caption{{\bf Flow chart of the algorithm.}
Flow chart showing the algorithm pipeline, including time series normalization, filtering, feature extraction, feature reduction, and data splitting into training and testing. The time signal shown here is one of the segments of the original time series. Note that the representation of blue bar for training dataset and green bar for testing dataset will be consistent in further discussions in this manuscript. The training data of all users were used for building $^{n}C_2$ binary classifier models, which becomes the process known as \textit{enrollment}.}
\label{fig:algorithm_pipeline}
\end{figure}

In addition to the use of MFDFA as a feature extraction algorithm, we also use of an automated time series feature extraction algorithm named \textit{tsfresh} (Time Series FeatuRe Extraction on the basis of Scalable Hypothesis tests) developed by \cite{CHRIST201872}.  The tool generates over $700$ time series features using 63 different time series characterization methods. The following discussion pertains to the preparation of dataset for model building, training and testing of the algorithms. A consolidated pipeline of the algorithm towards model library building including time series normalization, and selection, followed by feature extraction and reduction, is shown in Fig \ref{fig:algorithm_pipeline}.

Features extracted by these algorithms from all available time series are concatenated and passed through a low-variance filter. This was done to remove those feature columns with a variance value below a given threshold, which in our case was $1\%$. The rationale behind applying this low variance filter was to eliminate features that exhibit very little variation across instances. Such low-variance features may not provide useful insights for classification tasks. Furthermore, highly correlated features were removed from the feature set. A correlation threshold of $80\%$ was chosen for this purpose. Removing features by these techniques reduce the dimensionality, simplifies the model, and potentially improves model performance by focusing on more informative features. All features which were derived from the absolute values of the time series, such as maximum/minimum values, quantile information, etc., were disregarded. For example, inclusion of mean value of a signal will bias the algorithms and allow them to classify on the basis of the mean values itself, which was undesired. It was observed that different human subjects were able to exhale in different velocity bands depending on their lung capacity. The filtered feature matrix thus obtained is a stack of vectors from each time series sample available, and it consisted of approximately $450$ time series features. This feature space is high dimensional and may contain redundant features that can be excluded. The reduced feature set will also reduce the computational complexity of the algorithms. We adopted a feature selection method using binary random forest classifiers. Binary classifiers were built on pairwise combinations of the users' feature datasets. The importance of the features can be quantified for every random forest binary classifier by estimating how much the random forest's performance would suffer if a given feature were to be eliminated. This impurity-based feature importance developed by \cite{Breiman2001} was used for picking the top features. The top $10$ most prevalent features among users were chosen as the feature space after computing the top $10$ features from each classifier. In the later sections of this manuscript, the methods used for model construction and the physical insights of these features will be described. The reduced feature matrix thus obtained contains features of all the users in the database. For each user, the dataset was split into training $(60\%)$ and test $(40\%)$ sets. It is important to note that this splitting was done after shuffling between groups of features corresponding to the $19$ time blocks for each subject. We know that there were $190$ time signals for each user in the database with each set of $19$ signals coming from a single recorded time series (see subsection titled \textit{Time series segmentation, normalization and selection}). Shuffling without grouping would result in the the same information being spread across the training and testing dataset, which was undesired. By doing this we made sure that out of $10$ exhaled breath samples, $6$ become part of training set and $4$ become part of the test set. The training feature set was used to build the model library and the test feature set was used for user confirmation/identification tests.

\subsection*{Building of model library} \label{sec:model_library}
We have formulated the multi-class classification problem into a series of binary classification problems. Several studies have explored the application of pairwise binary classifiers for handling multi-class problems. A description of class binarisation and round robin classification can be found in \cite{Johannes2002}. \cite{Lorena2008}, \cite{Galar2011}, and \cite{Lan2022} are few others who have studied class binarisation for multi-class classification. In order to perform tests with a machine learning based algorithm, it was necessary to build binary classifier models using binary combinations of the training datasets and these models were stored in a \textit{model library}. Computational simulations were setup to evaluate the performance of the user confirmation and identification algorithms. Let us briefly see how the model library grows with the addition of users to the existing database of users. This is known as \textit{enrollment mode} of the biometric system. Say, there are $n$ disjointed users $U_1, U_2,\ldots U_n$ in the current state of the users' database. $^{n}C_2$ binary classifier models can be built, which makes up the complete model library. With the addition of a user, the updated size of the users' database becomes $n+1$. Therefore, the size of the model library increases by $n$ and becomes $^{n+1}C_2$. This growth can be expressed as
\begin{equation}
    ^{n+1}C_2 =\, ^{n}C_2 + n
\end{equation}
This means that when a new user is added to the users' database, $n$ additional binary classifier models are to be built and stored in the model library. Expectedly, this follows a second-order power-law variation of the form $y=ax^m$ with the multiplication factor $a\approx 0.5$ and exponent $m\approx 2$.

\section*{User confirmation algorithms} \label{sec:UCA}
Two different user confirmation algorithms were built using the extracted feature data. The first approach was based on statistical hypothesis testing, which involves the testing of a null hypothesis against an alternative hypothesis. The second approach was based on machine learning models. In case of a machine learning based algorithm development, the training data were used to build random forest binary classifier models, thereby creating a library of models. In the case of the hypothesis testing based algorithm, model building process is redundant, and the predictions are made based on the hypothesis test results between a user's test data and available training data, making it an \textit{instance-based} algorithm. These algorithms will be referred to as UCA.HT (User Confirmation Algorithm - Hypothesis Testing) and UCA.ML (User Confirmation Algorithm - Machine Learning) in later sections. The Hotelling's T$^2$ test (\cite{hotelling1931}) was used in UCA.HT, which is a multidimensional version of the Student's $t$-test.

\subsection*{Confirmation algorithm based on hypothesis testing}
\begin{figure}[!h]
\includegraphics[scale=0.165]{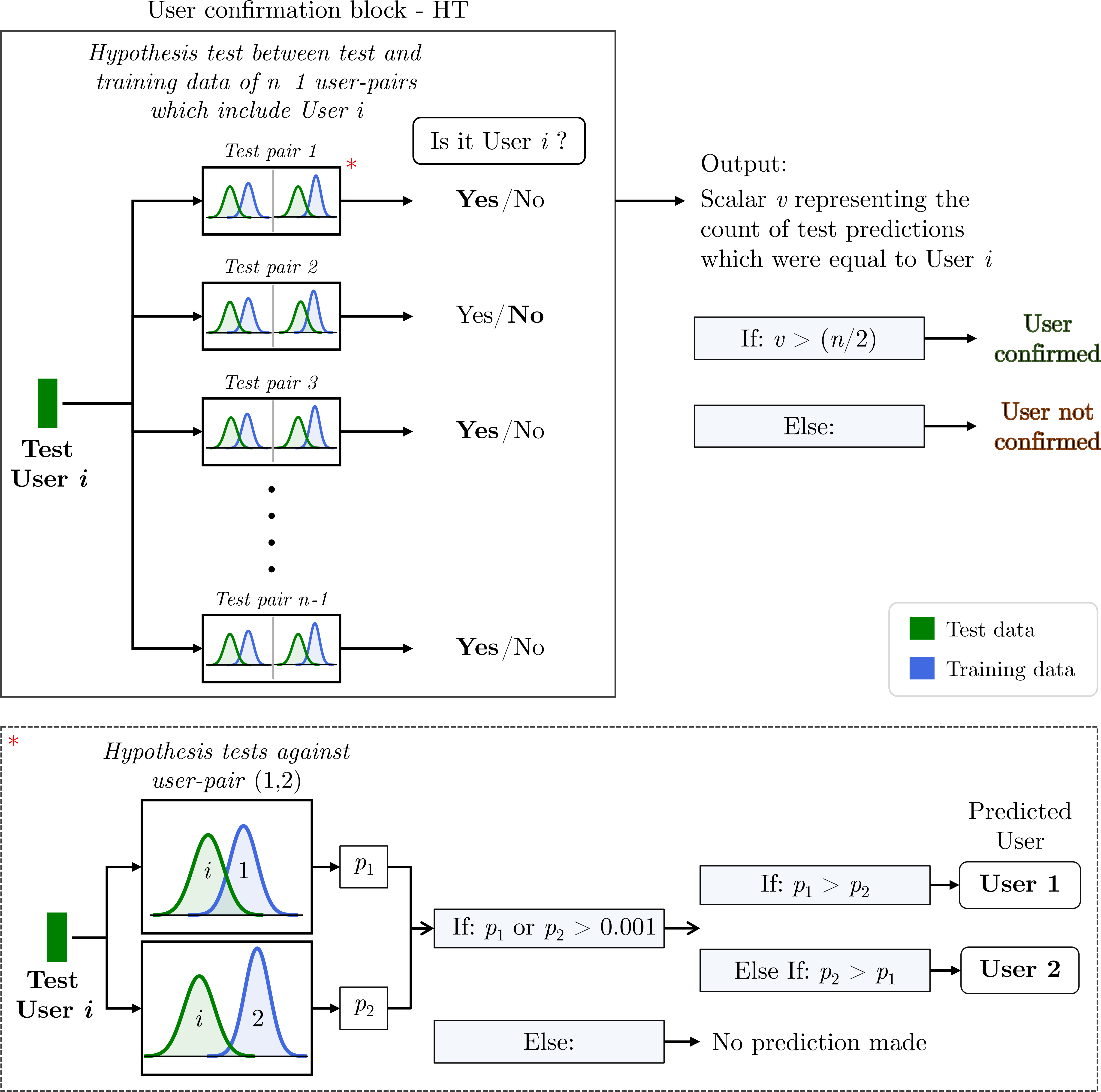}
\caption{{\bf User confirmation algorithm based on hypothesis testing.}
A flow chart of the user confirmation algorithm based on hypothesis testing. The user confirmation block will be made use in the user identification algorithm later in this manuscript. An example of the hypothesis test against user-pair is illustrated inside the dotted box, directed from the user confirmation block by the red asterisk. Given a user $i$, the user confirmation block's output was reposed to answer the question ``Are you indeed User $i$?'' based on a threshold.}
\label{fig:UCA_HT}
\end{figure}

The use of hypothesis testing as an instance-based binary classifier has been attempted in the literature. \cite{Janczura2020} compared the machine learning approach and the statistical testing based on $p-$variations; and the idea of instance-based classification by hypothesis testing was investigated by \cite{Zengyou2021}. \cite{Li2020} provided a detailed description on how binary decision problems can be formulated as hypothesis testing and/or binary classification. In a system based on hypothesis testing, the library comprises of the training datasets of all the users. Since we are building an algorithm which is intended to work alongside a machine learning algorithm, we formulate the hypothesis test based algorithm to work on binary pairs of users. To be more precise, the library will comprise of training datasets of pairs of users. It will be referred to as user-pair data in further discussions. Fig \ref{fig:UCA_HT} shows a flow chart of the user confirmation algorithm which is based on hypothesis testing principles. The equality-of-means test was performed between a test data and each training data in pairs present in the library to infer whether the null hypothesis is to be rejected or not, as depicted in Fig \ref{fig:UCA_HT}. Here, the null hypothesis states that the two samples come from the same distribution $(H_0: \mu_a=\mu_b)$, and the alternate hypothesis states that the samples come from different distributions $(H_1: \mu_a\neq\mu_b)$. A detailed description on the test statistic and formulation of the Hotelling's T$^2$ test can be found in the original work by \cite{hotelling1931}.

When a test user, say `User $i$' was to provide the input, the pairwise Hotelling's T$^2$ tests are performed between the test user's data and the training data of $n-1$ pairs of users which include `User $i$', where $n$ is the number of users in the database. Let us look at one of those tests as shown inside dotted box in Fig \ref{fig:UCA_HT}. By performing a hypothesis test against a user-pair, for example, $(1,2)$, we get a pair of $p-$values, $(p_1,p_2)$. The tests were performed with a confidence level of $99.9\%$, and therefore a $p-$value of $0.001$ or less was sufficient to reject the null hypothesis. At least one of the two $p-$values need to be above $0.001$ for the algorithm to accept the null hypothesis. The predicted user is then the user corresponding to a higher $p-$value. If both $p-$values are either equal to or below $0.001$, no predictions were made. After the test, the predictions made here are reposed as an answer to the question ``Is it User $i$? (Yes/No)''. The pipeline discussed so far becomes the `User Confirmation Block - HT' for the hypothesis testing based algorithm. The output of this block is a scalar $v$ which is equal to the count of model predictions which says `Yes'. Here, a threshold of $50\%$ of the predictions was used for defining the minimum confidence of confirmation. This means that HT$(i,i)$ accepts the null hypothesis and HT$(i,j)$ $\forall j=1,2,\ldots n$ and $i \ne j$ rejects the null hypothesis in at least $50\%$ of the cases. Then, the User $i$ is so confirmed. Here, HT$(i,j)$ stands for hypothesis test between a User $i$ and User $j$.

The equality-of-means test can actually be viewed from two perspectives: (a) Testing the distribution of test data against the distribution of $n$ training data; (b) Testing the distribution of test data against the distribution of training data in pairs as discussed so far. The former strategy produces $n$ test results and the algorithm would face one of three scenarios: $(i)$ If only one test accepts the null hypothesis, the user identity is presumed to be of the user corresponding to that particular test; $(ii)$ If more than one tests accept the null hypothesis, the user corresponding to the test which corresponds to highest $p$-value is presumed to be of the user identity (predicted user). In either case, if the predicted user matches with the test user, the user is confirmed, otherwise not; $(iii)$ If all tests reject or no test rejects the null hypothesis, then the user is not confirmed. Although the former case (procedure (a)) is a computationally simpler formulation, the latter case (procedure (b)) becomes more relevant in our study since we are trying to build a multi-model approach for user identification. It was also noted that the latter approach gave better confirmation results (for UCA.HT) compared to the former approach.

\subsection*{Confirmation algorithm based on machine learning} \label{sec:uca.ml}
Following the discussions from subsection titled \textit{Building of model library}, generating $^{n}C_2$ binary classifiers is necessary to handle the multiclass problem. Let us have a detailed discussion on the model building procedure and the choice of the binary classifier. We required a detailed analysis since the choice of a classifier depends on the specific characteristics of the dataset and the multiclass problem at hand. The training dataset was used to construct binary classifier models for each user-pair. Decision tree (DT), random forest (RF), support vector machine (SVM), logistic regression (LR), Gaussian naive Bayes (GNB), and multi-layer perceptron (MLP) were chosen as the candidate binary classifier models. The machine learning models employed in our study are discriminative models except for one, the GNB. Discriminative models do not try to identify the distribution that generates the data; instead, they try to find out the features that separate classes from each other (\cite{Ng2001}). 

\begin{figure}[!h]
\includegraphics[scale=0.14]{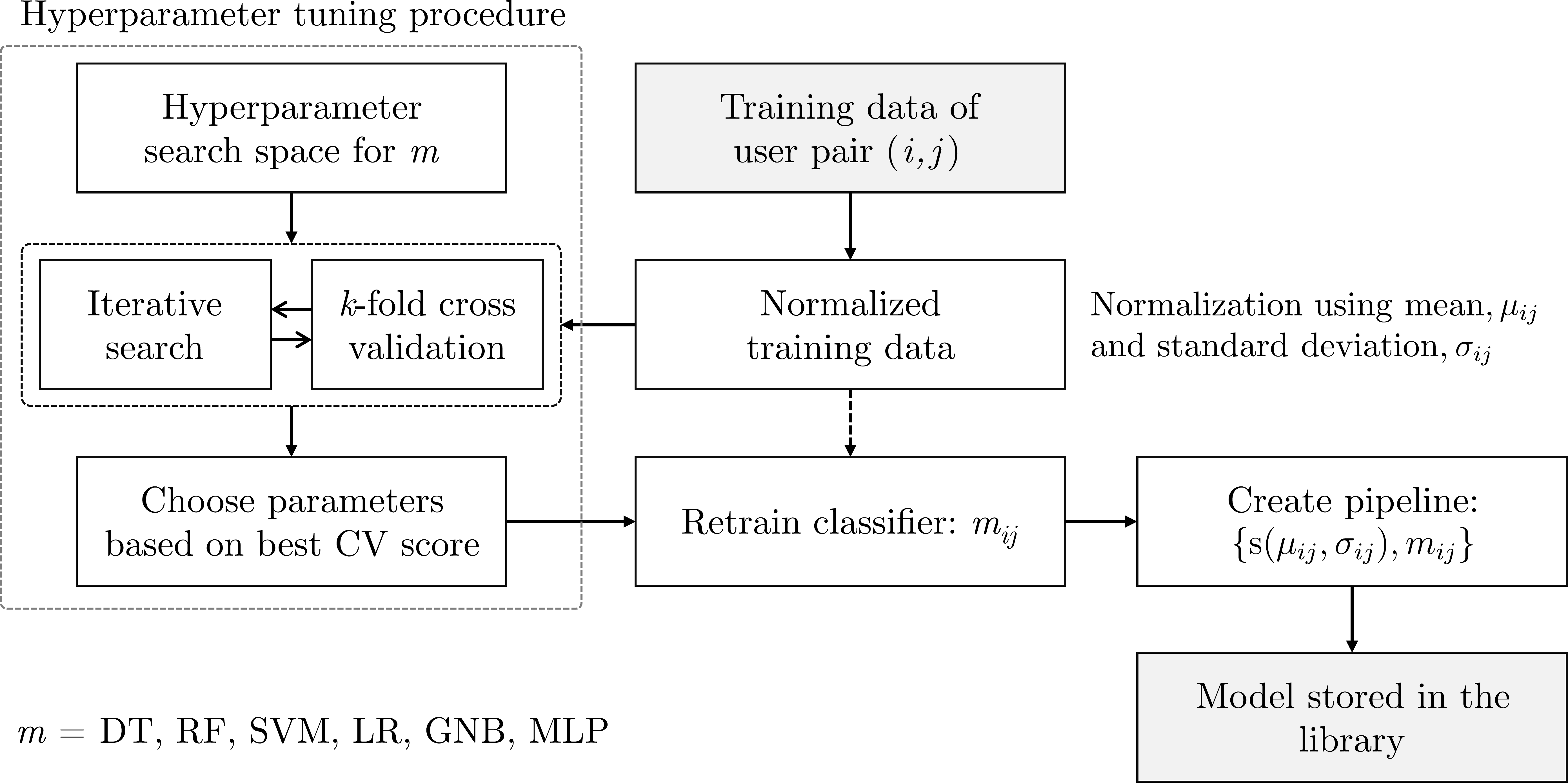}
\caption{{\bf Model library building procedure.}
Flow chart showing the model library building procedure for a user-pair $(i,j)$, where $i=1,2,\ldots n$; $j=1,2,\ldots n$; $n$ is the total number of users. Note that model $m_{ij}\equiv m_{ji}$, and therefore, only model $m_{ij}$ are built and stored in the library. The abbreviations stand for the following: CV - cross validation, DT - decision tree, RF - random forest, SVM - support vector machine, LR - logistic regression, GNB - Gaussian naive Bayes, MLP - multi-layer perceptron. $\mathrm{s}(\mu_{ij},\sigma_{ij})$ is the standard scaling function, $\mu_{ij}$ and $\sigma_{ij}$ are the mean and standard deviation respectively of the training data of users $i$ and $j$ combined.}
\label{fig:model_building}
\end{figure}

Robustness of model parameters and selection of the best model is very crucial to the performance of a user authentication algorithm. Optimal tuning improves the generalizability of each machine learning model. A generic algorithm for hyperparameter tuning and model selection is illustrated in Fig \ref{fig:model_building}. The training data for a user-pair $(i,j)$ is normalised initially using the training set's mean $(\mu_{ij})$ and standard deviation $(\sigma_{ij})$. $\mu_{ij}$ and $\sigma_{ij}$ should be stored in the memory as it is required for scaling the test data when required. Hence, it can be combined into a function called standard scaling function $\mathrm{s}(\mu_{ij},\sigma_{ij})$ for later use. This normalised training data is now used for tuning and training the best model. It is generally challenging to know the values of the model parameters for a given machine learning model on a dataset. Therefore, we have employed an iterative search cross-validation scheme to compare different sets of hyperparameter values for each model. A stratified $k$-fold cross-validation technique with hyperparameter tuning was employed for evaluation and selection of the model parameters. The number $(k)$ of folds was chosen to be $5$. Parameters from a hyperparameter search space were fed into the cross-validation algorithm, where the training data was split into $k$ equally sized folds maintaining the same target class distribution in each fold as the original dataset. This will make sure that there is no class imbalance in each of the $k$ folds. The iterative search from the search space was performed either using Bayesian optimisation based search (\cite{Bergstra2011}, \cite{Bergstra2013}) or by grid search depending on the size of the search space of a classifier model. The Bayesian search method employs a probabilistic model of the search space to choose a hyperparameter configuration. By combining exploration (experimenting with new configurations) and exploitation (utilizing knowledge from previous iterations), it effectively explores the hyperparameter space and identifies promising regions in the hyperparameter space, as described by \cite{Snoek2012}. The grid search can be described as an exhaustive exploration method which tests all the combinations of the search space (\cite{Belete2022}). Bayesian search was employed when the search space was large, whereas the grid search was employed for smaller search space where the method of brute force was computationally affordable. During an iteration, for a selected hyperparameter configuration, the model was trained and evaluated $k$ times, each time using a different fold as the validation set and the remaining $k-1$ folds as the training set. The set of parameters which yielded the best cross-validation score were selected, and the model was retrained on the normalised training data based on the selected parameters. The standard scaling function and the classifier model are combined into a model pipeline $\left\{\mathrm{s}(\mu_{ij},\sigma_{ij}), m_{ij}\right\}$, where $m_{ij}$ is the classifier built corresponding to user-pair $(i,j)$. The pipeline was create to make sure that the test data when introduced should be scaled using the training set's mean and standard deviation before predictions are made.

The above procedure was performed for all the candidate binary classifiers. This procedure helped in effectively reducing the size of the feature set from $\approx 450$ dimensions to $10$ dimensions using random forest classifiers as briefed in subsection titled \textit{Feature extraction}. The number of trees/estimators were tuned. More trees are generally required for the purpose of refined variable importance estimations, as noted by \cite{GENUER20102225}. The rule of splitting was tuned by controlling the maximum depth of a tree, minimum number of samples required to split an internal node, and minimum number of samples required to be at a leaf node. It was important to Fig out the best model based on their performance. This will allow us to build an efficient library of best estimators for the given training data. Each of the $^{n}C_2$ models underwent a hyperparameter tuning before being fit to the training data. This way it was ensured that each model is generalized for the corresponding user-pair's data. It is important to note here than any model with a cross-validation score of $60\%$ or less was discarded. This was done to ensure that the models which were saved in the library are far from a random model. Hence, by storing only the models with a cross-validation score above $60\%$, we make sure that the overall algorithm performs between reasonably well to good, based on the generalisation of the models built.

\begin{figure}[!h]
\includegraphics[scale=0.455]{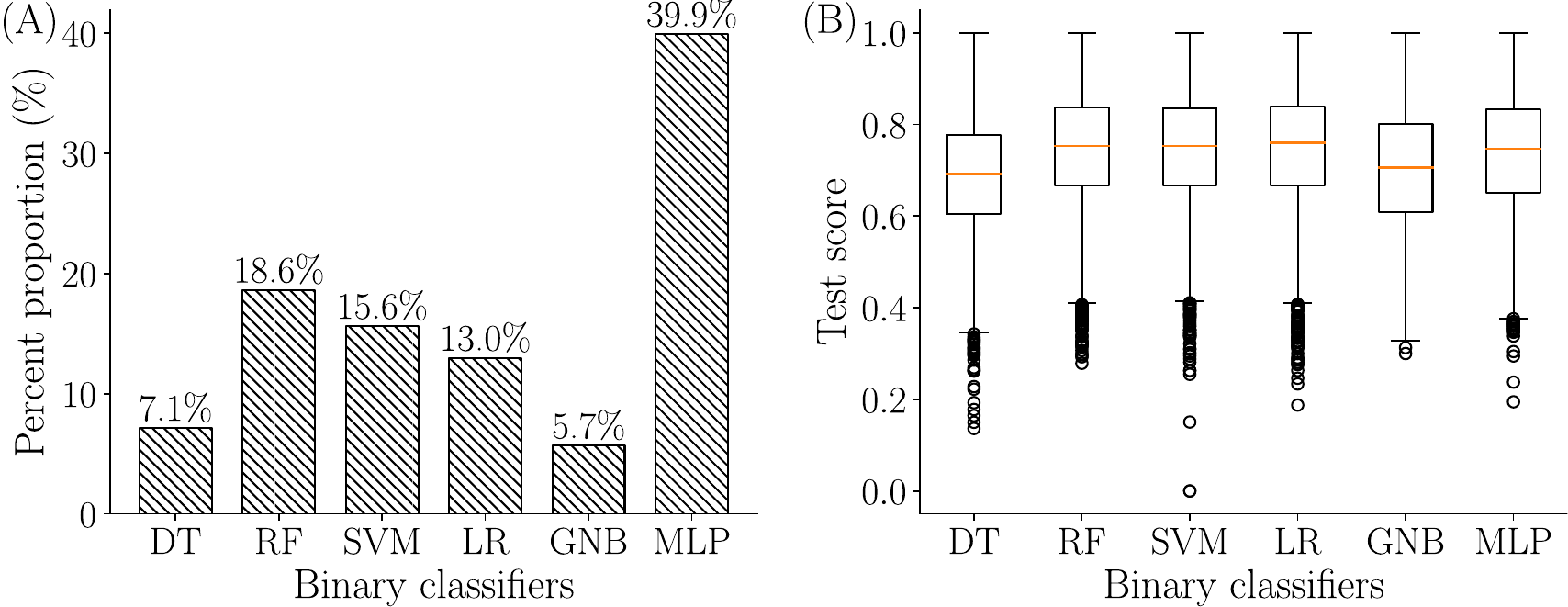}
\caption{{\bf Comparison of candidate classifier models.}
(A) Bar chart showing the percent proportion of each model in the library in the case of \textit{best-of-all} model selection procedure. (B) Box and whiskers plot showing the spread of test accuracy of each classifier. The orange line inside the boxes represent the median.}
\label{fig:model_choice}
\end{figure}

Selection of the best binary classifier model for a given user-pair can be made by picking the model with the highest cross-validation score. This technique can be called as \textit{best-of-all} model selection technique. Fig \ref{fig:model_choice}A shows the percent proportion of different models in a library which was built using this procedure. We observed that MLPs were the most frequently occurring best classifier based on the highest cross-validation score. The second frequently occurring one being the RF, followed by SVM, LR, DT and GNB. The process of building all $6$ models and choosing the best one every time is computationally very expensive, So, one out of the $6$ classifiers had to be chosen for testing the algorithms. The information from Fig \ref{fig:model_choice}A is not sufficient to make this decision, since a good cross-validation score does not always promise a good performance on test data since the classifiers could also be overfitting the training dataset in certain cases. Pairwise user test data was used to test each candidate classifier and the results are visualised as a box and whiskers plot in Fig \ref{fig:model_choice}B. The orange line inside the boxes represents the median of the test score. It it clear that DT and GNB classifiers perform poorer than the others, and, RF, SVM, LR and MLP have very similar performance on the test data. All the models have produced test accuracies ranging from very low values below $0.5$ to $1$. Looking at the outliers data points, we can see that SVM does very poorly as the accuracy even goes below $0.2$, and in fact LR and MLP too have produced accuracies below $0.2$. The RF classifiers and GNB have similar lower bound of test accuracy.

\begin{figure}[!h]
\includegraphics[scale=0.2]{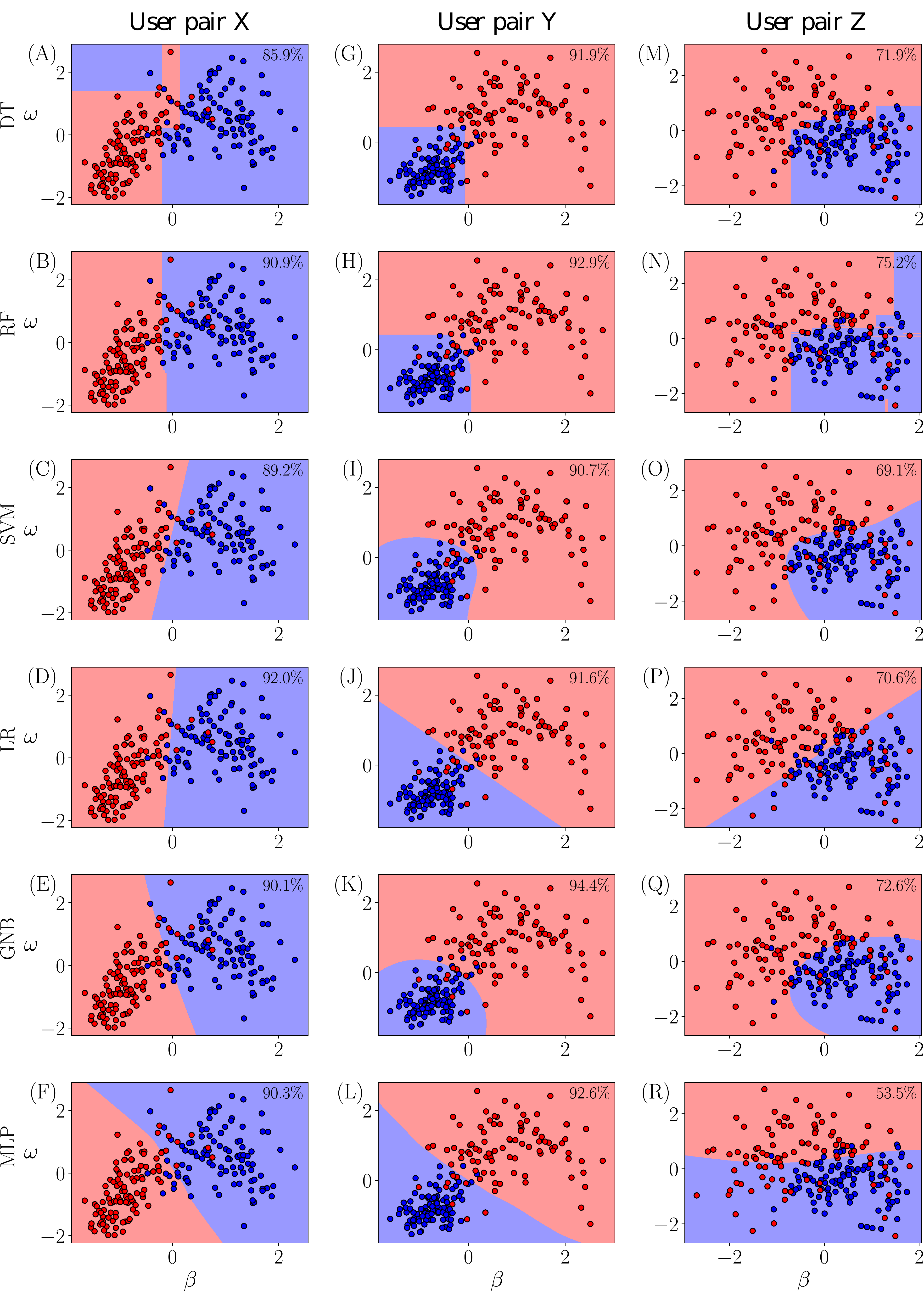}
\caption{{\bf Two dimensional decision boundaries.}
Comparison of two dimensional decision boundaries in the $(\beta, \omega)$ plane, captured by different models for three randomly chosen user-pairs. The scattered points are the training data points with red and blue labels denoting their true classes respectively. The line separating the two contour regions is the decision boundary. Accuracy of each model against the test data is displayed at the top right corner of their respective plots. The abbreviations stand for the following: DT - decision tree, RF - random forest, SVM - support vector machine, LR - logistic regression, GNB - Gaussian naive Bayes, MLP - multi-layer perceptron.}
\label{fig:decision_boundary_training}
\end{figure}

In order to get a better understanding on how these models fit the training data, we can visualise the decision boundaries in a 2D feature space. Since we are already working on a reduced feature space of $10$ dimensions, choosing any $2$ dimensions out of it and building models for the purpose of visualisation seems appropriate here. The $(\beta,\omega)$ space was chosen for visualisation. To generate the 2D decision boundaries, a structured synthetic dataset was generated which filled up the two-dimensional feature space within the given bounds. The decision regions are obtained based on the predictions made on each data point from the synthetic dataset. These boundaries are visualised for comparison in Figs \ref{fig:decision_boundary_training}A$-$\ref{fig:decision_boundary_training}R for three randomly chosen user-pairs.

\newpage

A region $\mathrm{R}$ in the feature space is classified as a decision region under class $y_i$ $(i=\{0,1\})$ if all the samples $x_j$ in that region is classified as $y_i$. A decision boundary separates these 2 decision regions. Therefore, it can be observed that the feature space is divided into two parts by the decision boundary for a binary classification problem. Such a representation not only helps us visualize the difference between two classes, but also helps us in comparing multiple binary classifiers and their decision mechanisms. The scattered points in each plot of Fig \ref{fig:decision_boundary_training} represent the training data points with their respective colors corresponding to two users. The test data accuracy for each model is displayed at the top right corner of their respective plots. Considering user-pair X (Figs \ref{fig:decision_boundary_training}A$-$\ref{fig:decision_boundary_training}F), all the models were able to perform well, with LR and RF producing the scores higher than the other choices. Similarly, for user-pair Y (Figs \ref{fig:decision_boundary_training}G$-$\ref{fig:decision_boundary_training}L), all models were able to perform well, with GNB, MLP and RF producing the best scores. It is interesting to see that the decision boundaries captured by SVM (Fig \ref{fig:decision_boundary_training}H) and RF (Fig \ref{fig:decision_boundary_training}K) appear similar but small variations in the captured boundary causes one algorithm to perform better. Considering user-pair Z (Figs \ref{fig:decision_boundary_training}M$-$\ref{fig:decision_boundary_training}R), the models produce less accuracy against the test data when compared with user-pairs X and Y. GNB and RF produce the highest scores among the models. From the discussion so far, it is evidenced that the random forest models (see Figs \ref{fig:decision_boundary_training}B, \ref{fig:decision_boundary_training}H and \ref{fig:decision_boundary_training}N) were able to capture a complex decision boundary in all the three cases and are able to perform well in all the three cases. This could be due to its bootstrapping and ensemble schemes, making it robust to outliers. Also, RF is known to reduce the risk of overfitting by aggregating predictions from multiple decision trees and generalising well. Therefore, we chose random forest as the apt binary classifier model for the model library. All the $^{n}C_2$ trained models (the pipelines, as discussed earlier) were stored in the library, where $n$ is the total number of users. For the rest of this work, we will employ random forest as our machine learning algorithm  and report results from this tool for both user confirmation and user identification.

\begin{figure}[!h]
\includegraphics[scale=0.165]{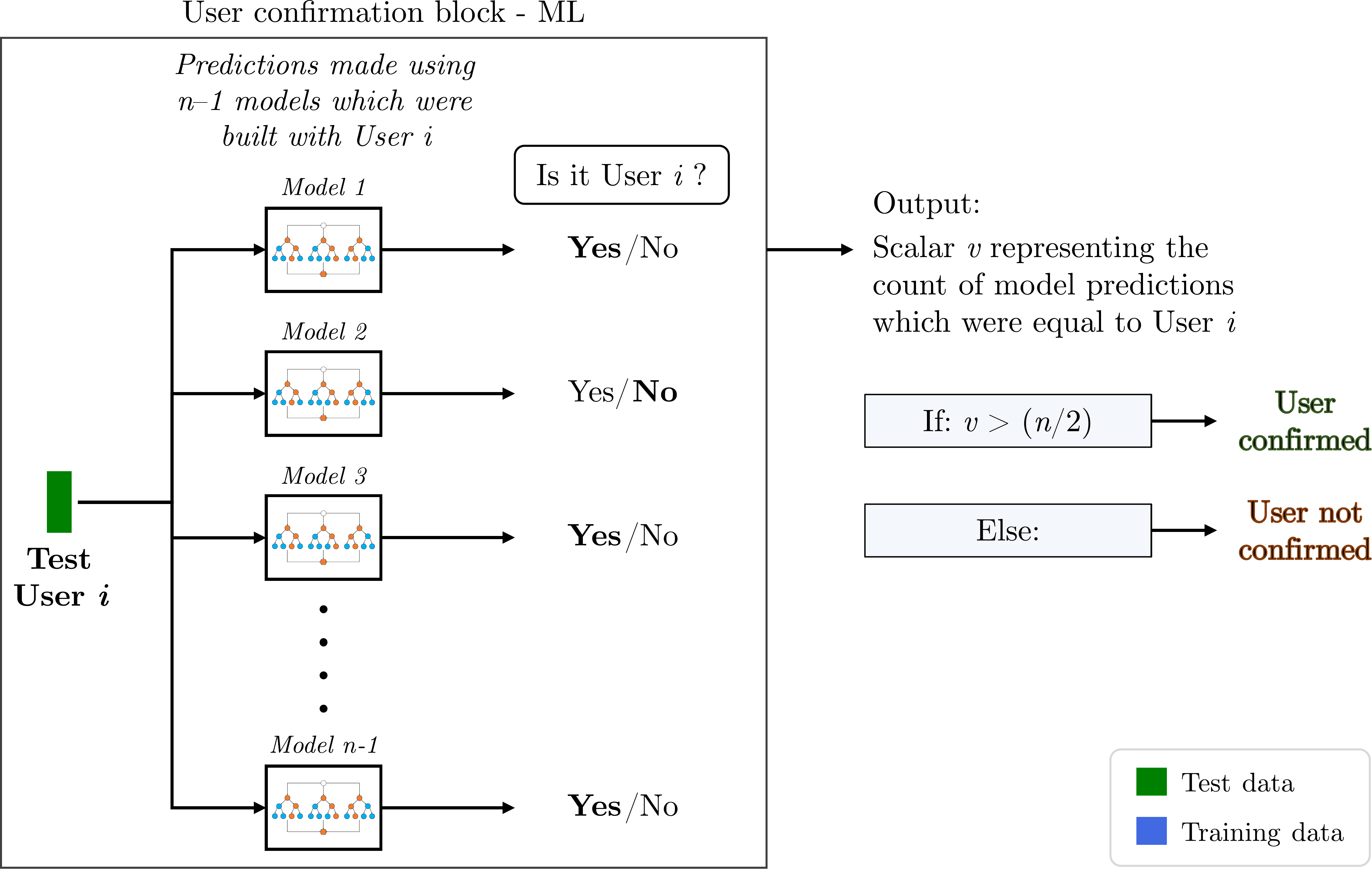}
\caption{{\bf User confirmation algorithm based on machine learning.}
A flow chart of the user confirmation algorithm based on machine learning. The user confirmation block will be made use in the user identification algorithm later in this manuscript. Given a user $i$, the user confirmation block's output was reposed to answer the question ``Are you indeed User $i$?'' based on a threshold.}
\label{fig:UCA_ML}
\end{figure}

Once the model building was complete and the entire library was stored, the test user data were given as input, say `User $i$'. The algorithm selects those models from the library which were built using the same test user and makes predictions using each model as depicted in the flow chart in Fig \ref{fig:UCA_ML}. The predictions made here are answers to the question ``Is it User $i$? (Yes/No)''. The pipeline discussed so far becomes the `User confirmation block - ML'. The output of this block is a scalar $v$ which is equal to the count of model predictions which says `Yes'. Here, a threshold of (again) $50\%$ of the predictions was used for defining the minimum confidence of confirmation. This means that if the algorithm confirms the user in more than half the classification trials, i.e., when $v>(n/2)$, the user is confirmed, else not.

\section*{User identification algorithm}
\begin{figure}[!h]
\includegraphics[scale=0.165]{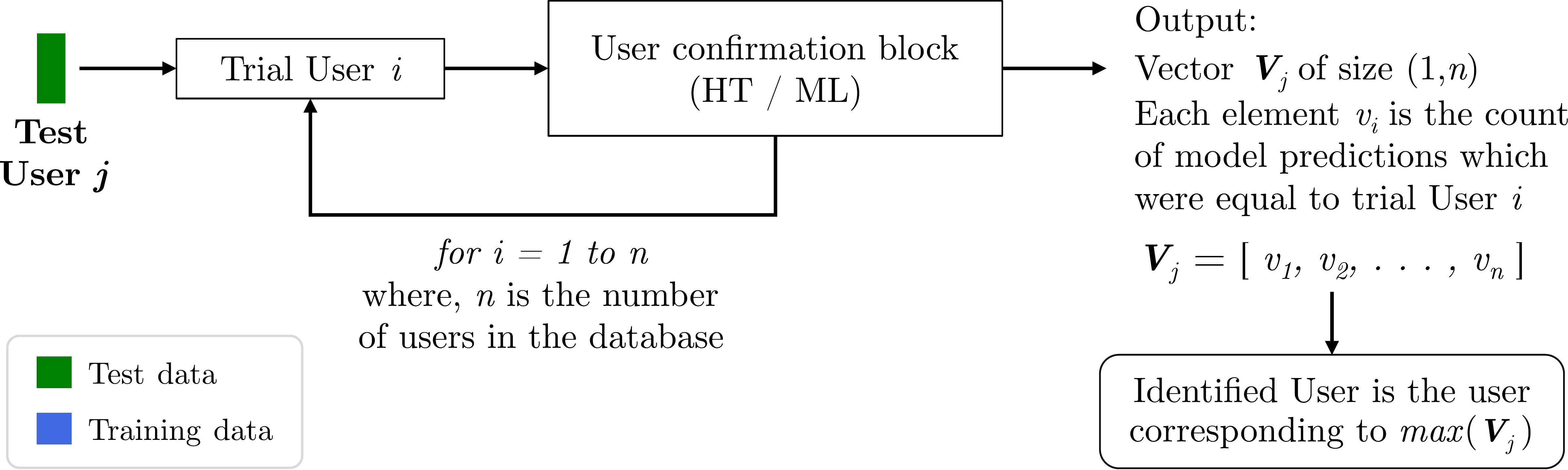}
\caption{{\bf A generic user identification algorithm.}
Given a test user $j$, the algorithm performs $n$ confirmation trials. One confirmation trial is the equivalent to running the user confirmation block (either HT from Fig \ref{fig:UCA_HT} or ML from Fig \ref{fig:UCA_ML}) for a trial user $i$. The identified user corresponds to the maximum prediction based on the $n$ confirmation tests. Note that in the case where more than one confirmation trial results in the maximum prediction value, the algorithm does not identify a user.}
\label{fig:UIA_ML}
\end{figure}

This work is the first attempt of its kind to build a biometric system which works purely based on human exhaled breath to identify the user with no disclosure of the user's identity by the user himself or herself. Even though the user confirmation system works exceptionally well, the grand challenge in this area of research is to test the performance of a user identification system. The confirmation algorithm tries to answer the question ``Are you User $i$?'', while an identification algorithm would essentially answer the more general authentication question ``Who is the User?''. In pursuit of this grand challenge, we have developed a user identification algorithm built on top of approaches discussed in this manuscript. The machine learning based algorithm would use the same model library built earlier to perform the predictions. A block diagram of the algorithm is shown in Fig \ref{fig:UIA_ML}. The user identification algorithm incorporates the user confirmation block during the identification of a given user. When a new test user data is given as input, say User $j$, the algorithm runs the user confirmation block by considering all the users in the database as trial users. This effectively is equal to running through all the $^{n}C_2$ models present in the library, but in batches of trial users, User $i$, where $i=1,2,3,\ldots n$. The output of this pipeline is a vector $\boldsymbol{V}$ of size $(1,n)$ with each element $v_i$ being a result of the corresponding trial confirmation test. The identified user from this algorithm will be the trial user corresponding to the maximum value in the vector $\boldsymbol{V}$. When more than one confirmation trial results in the maximum prediction value (two elements of $\boldsymbol{V}$ having the maximum value), the algorithm does not identify any user. The user identification algorithm is made generic, which means that any user confirmation algorithm (instance-based or model-based) can be used within this algorithm and the output of this algorithm will be the vector $\boldsymbol{V}$ containing the count of predictions. This allows us to build a multi-modal approach for user identification where multiple identification results can be combined using a weighted sum. This is similar to a classical black board architecture where results from multiple expert units are combined. We will now present a brief discussion on this approach. Let us call the outputs from a hypothesis test based and machine learning based user identification algorithms as $\boldsymbol{V}^{\text{HT}}$ and $\boldsymbol{V}^{\text{ML}}$ respectively. We can take a weighted sum of these two vectors to get a new vector $\boldsymbol{V}^{'}$ which will have the advantages of both the algorithms as shown in Eq \ref{eq:wV}.

\begin{equation} \label{eq:wV}
    \boldsymbol{V}^{'} = w_1 \times \boldsymbol{V}^{\text{HT}} + w_2 \times \boldsymbol{V}^{\text{ML}}
\end{equation}
where, $w_1$ and $w_2$ are the weights associated with hypothesis test based algorithm and the machine learning based algorithm, respectively. The weights can take values between $0 \le w \le 1$ and sum of the weights should always sum up to $1$. This approach can be generalised for a combination of multiple user identification algorithms as shown in Eq \ref{eq:weighted_sum}. When we have $r$ output vectors $\boldsymbol{V}_1, \boldsymbol{V}_2, \boldsymbol{V}_3, \ldots, \boldsymbol{V}_r$ from $r$ algorithms, Eq \ref{eq:wV} becomes,

\begin{equation} \label{eq:weighted_sum}
\boldsymbol{V}^{'} = w_1 \times \boldsymbol{V}_1 + w_2 \times \boldsymbol{V}_2 + w_3 \times \boldsymbol{V}_3 + \ldots + w_r \times \boldsymbol{V}_r
\end{equation}

\section*{Results and discussions}
\subsection*{User confirmation system}
\begin{figure}[!h]
\includegraphics[scale=0.45]{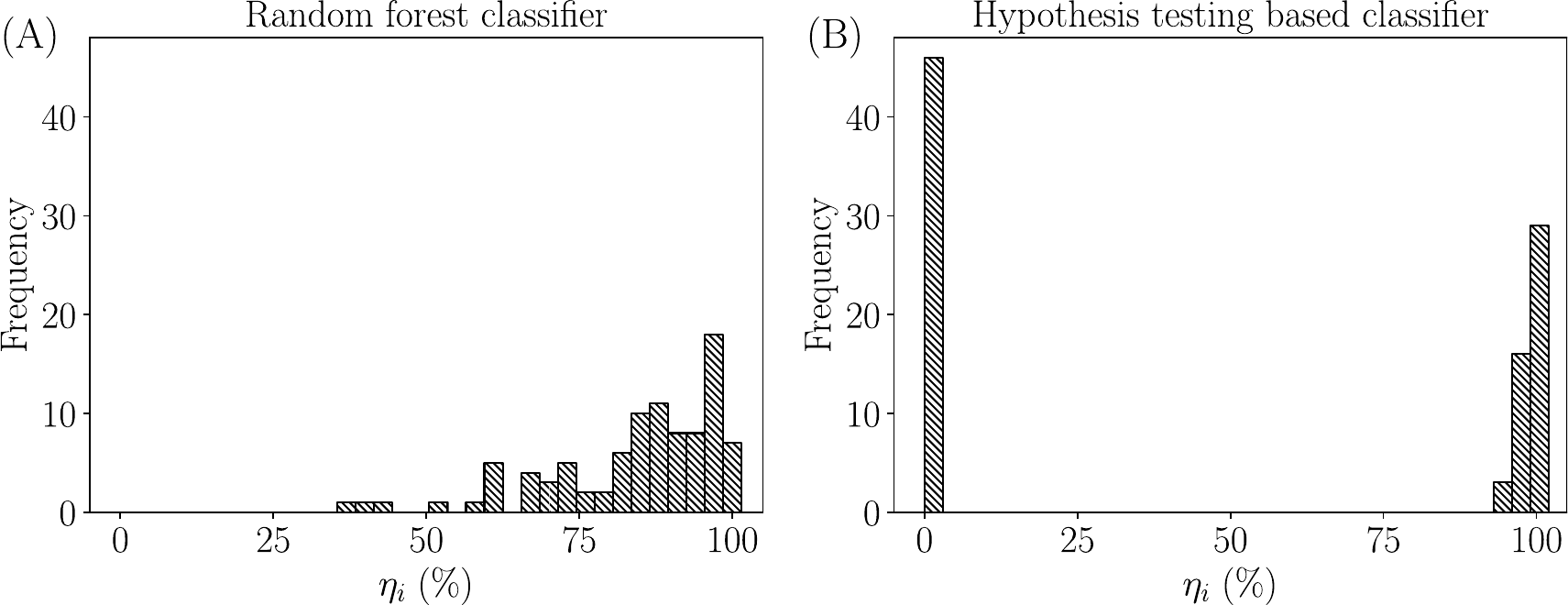}
\caption{{\bf Comparison of the confidence of confirmation $\eta_i$.}
Histograms of confidence of confirmation $\eta_i$ compared between (A) a machine learning based approach (random forest classifiers) and (B) a hypothesis testing based classification approach, for one trial of $n$ confirmation tests. In the example shown here, the predictions from ML classifiers give a range of $\eta_i$ values distributed between $\approx38\%$ to $100\%$, whereas the predictions from HT based classifiers produce $\eta_i$ values only close to $0\%$ and $100\%$.}
\label{fig:confidence_histogram}
\end{figure}

Confirmation tests were performed for all users $(n)$ available in the database. Each set of confirmation tests were repeated $66$ times by shuffling training and test data split-up. The results of the algorithm from each of these trials can be interpreted as follows: number of confirmed users denoted by $c$, and number of unconfirmed users denoted by $u$. In order to quantify the performance of the algorithms, we define a metric called the true confirmation rate (TCR) which is a ratio of the confirmed users and total number of users as shown in Eq \ref{eq:tcr}.
\begin{equation} \label{eq:tcr}
    \text{TCR}=\frac{c}{n} \times 100
\end{equation}
\noindent
The confidence of confirmation $(\eta)$ for a user confirmation algorithm is the percentage prediction of the favourable user during a confirmation test. It directly quantifies how confident the algorithm is while attempting to confirm a user $i$. It can be defined as,
\begin{equation} \label{eq:eta}
    \eta_i=\frac{v_i}{n-1} \times 100
\end{equation}
where, $v_i$ is the favourable user predictions as seen in Figs \ref{fig:UCA_HT} and \ref{fig:UCA_ML}, i.e., the total number of model predictions that matches the user that the algorithm is attempting to confirm, and $n$ is the total number of users in the database. The value of $\eta_i$ $\forall i=1,2,\ldots n$ has to pass a threshold confidence of confirmation, say $\eta_t$, for a user to be confirmed. A comparison of the histogram of $\eta_i$ is shown in Fig \ref{fig:confidence_histogram} for one trial of $n$ confirmation tests. The study revealed that the machine learning based algorithm performs better than the hypothesis testing based algorithm. This validates the ability of a random forest classifier to capture the decision boundary better, when compared to its hypothesis testing based counterpart. For the UCA.HT, the TCR was $50 \pm 9.6\%$, whereas, for the UCA.ML, the TCR was $97 \pm 2.5\%$. This implies that almost every user was able to pass the threshold of $50\%$ in the machine learning based algorithm. This signifies that the algorithm achieves a greater level of confidence while confirming a user using UCS.ML.

\begin{figure}[!h]
\includegraphics[scale=0.45]{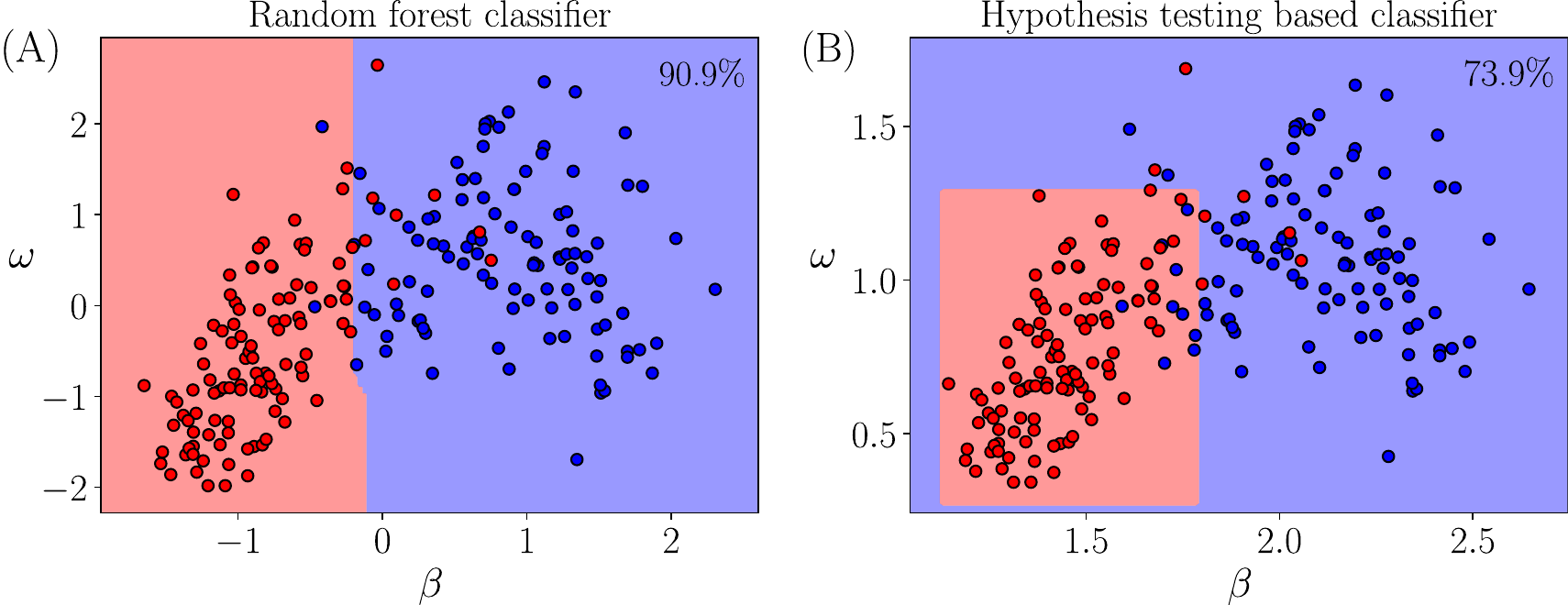}
\caption{{\bf Comparison of the decision boundaries in $(\beta,\omega)$ plane.}
Decision boundaries captured by (A) random forest classifier and (B) hypothesis testing based classifier for a randomly chosen user-pair. The scattered points are the training data points with red and blue labels denoting their true classes respectively. The line separating the two contour regions is the decision boundary. Accuracy of each model against the test data is displayed at the top right corner of their respective plots. The RF classifier captures a complex decision boundary compared to the HT based classifier.}
\label{fig:decision_boundary_rf_ht}
\end{figure}

We shall now investigate why the machine learning based classification algorithm performs better in comparison with a hypothesis test based classification. In the case of hypothesis testing, we know that the rejection of null hypothesis is based on the confidence level chosen. The confidence level can be visualized as a demarcating hyper-surface between two n-dimensional normal distributions. For simplicity, let us have a look at the decision boundaries captured by the random forest classifier and the hypothesis test based classifier in a chosen two dimensional feature space. Fig \ref{fig:decision_boundary_rf_ht} shows a visualisation in the $(\beta,\omega)$ plane for a randomly chosen user-pair. The blue and red markers are the training data points corresponding to two user classes, respectively. The class regions are computed using a structured synthetic dataset in the feature space.

For the purpose of visualisation of a hypothesis test based classifier's decision boundary, $z-$tests were performed in each dimension separately, for every data point from the synthetic dataset against one of the user's training data. The tests were performed under the null hypothesis that the data point belongs to the distribution of the training data, under a confidence level of $99.9\%$. The overall null hypothesis is accepted only if the null hypothesis in both the dimensions are accepted. Comparing the decision boundaries captured by a hypothesis test based algorithm and a random forest model for the same pair of users, one can observe that the random forest model has the ability to capture a more complex decision boundary between two user classes. This lets the random forest classifier to achieve a test data accuracy of $90.9\%$, whereas the hypothesis testing based classifier achieves only $73.9\%$. Now that we have established that the machine learning based algorithm is better than the hypothesis test based algorithm for user confirmation, we will now investigate how these two algorithms perform for user identification in the following section.

\subsection*{User identification system}
The identification algorithm discussed in Fig \ref{fig:UIA_ML} shows that we obtain a vector $\boldsymbol{V}$ of favourable user predictions. Based on the values of vector $\boldsymbol{V}_j$ with $j=1,2,3,\ldots n$, we can obtain the following outcomes:
\begin{itemize}
    \item True positives $(t)$ - Number of users who were identified correctly.
    \item False positives $(f)$ - Number of users who were identified incorrectly.
    \item Not identified $(h)$ - Number of users who the algorithm was unable to identify.
\end{itemize}
We shall define the following performance metrics to evaluate the user identification algorithm:
\begin{enumerate}
    \item Precision $(\mathrm{P})$ or Positive Predictive Value (PPV), which quantifies the percentage of users who were identified correctly among all the identified users.
    \begin{equation} \label{eq:precision}
        \mathrm{P}=\frac{t}{t+f} \times 100
    \end{equation}
    This parameter quantifies the probability of correct predictions given a judgement (identification) by the algorithm.
    \item Accuracy $(\mathrm{E})$, which quantifies the percentage of users who were identified correctly among all the users $n$.
    \begin{equation} \label{eq:accuracy}
        \mathrm{E}=\frac{t}{n} \times 100
    \end{equation}
\end{enumerate}

The precision and accuracy values computed using Eq \ref{eq:precision} and Eq \ref{eq:accuracy} respectively, were $35 \pm 10.5 \%$ and $29 \pm 9.1 \%$ respectively, for the hypothesis test based algorithm. The results reported in this section are in the format `$\mu_p\pm2\sigma_p$' where $\mu_p$ and $\sigma_p$ are mean and standard deviation of the performance metrics respectively. For the random forest based algorithm, we were able to observe precision and accuracy values of $26 \pm 7.2 \%$ and $22 \pm 6.4 \%$ respectively. These values were computed on the basis of the maximum votes received by a user among $n$ confirmation trials, as described previously in Fig \ref{fig:UIA_ML}. When we combine the results from both the algorithms using Eq \ref{eq:wV} with $w_1=0.3$ and $w_2=0.7$, we get precision and accuracy values of $32 \pm 8.5 \%$ and $31 \pm 8.5 \%$ respectively. Note that the values reported here are also influenced by the threshold $\eta_t$ which in this case was set to $55\%$. The parameters $w_1$, $w_2$, and $\eta_t$ can be tweaked to make the algorithm behave on both extremes - $(i)$ to be very liberal (low precision, low accuracy); $(ii)$ to be very conservative (high precision, low accuracy). Taking the example of a particular trial with $n=94$, for a weights setting of $w_1=0.3$ and $w_2=0.7$, $\eta_t=50\%$ produces the outcomes $(t,f,h)=(31, 58, 5)$, giving a precision of $34.8\%$ and accuracy of $33.0\%$. For the same weights, $\eta_t=96\%$ produces the outcomes $(t,f,h)=(18,6,70)$, giving a precision of $75.0\%$ and accuracy of $19.1\%$. The former case allows for a lot of false positives by making judgements on most of the instances, whereas the latter case of the algorithm makes judgements stringently.

With the right set of hyperparameters ($w_1$, $w_2$, $\ldots w_r$ (in the general case from Eq \ref{eq:weighted_sum}) and $\eta_t$), a multi-modal approach is expected to improve the robustness of the overall algorithm. If one classifier produces incorrect predictions for certain trials, other classifiers in the ensemble can compensate for it and provide correct predictions. The contribution of each algorithm can be controlled by the weights. This robustness helps in improving the generalization of the ensemble model. The following discussion is based on results produced from this combined algorithm. We know that the highest voted user becomes the identified user from the algorithm. Based on the $66$ shuffle trials, we have the following understanding of the user database. $21.3\%$ to $42.6\%$ of the users can be correctly identified by them being the highest voted users, $39.4\%$ to $57.4\%$ of the users can be correctly identified as at least the second highest voted users, and $50.0\%$ to $66.0\%$ of the users can be correctly identified as at least the third highest voted users. This is remarkable given that it is the first attempt in the literature to classify and uniquely identify individuals based solely on the fluid physics of the exhaled breath. We believe that this is conclusive evidence that the fluid dynamic structure of the exhaled breath contains uniquely identifiable information.

This algorithm holds tremendous potential for future use in the area of personalised medicine and also as a novel way to store biological data. This can be achieved by careful model selection and generalisation of classifier models. Advanced models such as deep neural networks can be made use to enhance the multi-model approach discussed in this manuscript.

\subsection*{Physical insights: Understanding the defining features}
In order to make a physics-based argument for the uniqueness of human exhalation, it is important to investigate the physical significance of the most important features that result in robust classification. These would be the set of features or attributes which inherently differentiate the classes for a given training data. As we have seen, the importance of the features were quantified for every random forest binary classifier for choosing a reduced feature set in subsection titled \textit{Feature extraction}. These features are to be investigated to understand their physical meaning in the context of the current problem in hand. A description of the most important classifying features (in the decreasing order of importance) are as follows.

\begin{enumerate}

    \item The singularity strength or Hölder exponent corresponding to the maximum ($\beta$) of the multifractal spectrum of the exhaled breath time series: This is a feature extracted using the MFDFA. $\beta$ explains the long range correlation present in the time series. A low value indicates that the underlying process becomes correlated and loses fine structure, becoming more regular in appearance (\cite{kantelhardt2002multifractal}). This, in our case, would relate to the organised motion of vortical structures in the turbulent exhaled air flow. For some subjects the vorticity pattern might be more irregular than the others, which could be attributed to the extrathoracic morphology.

    \item The sum over the absolute value of consecutive changes in the velocity time series: This is a measure of similarity between consecutive time blocks, and this metric describes the existence of mean reverse if present. Such a characteristic can infer the existence of vortical structures in the exhaled flow being unique for every individual, allowing them to be classified by the algorithm.

    \item Third coefficient of the autoregressive $AR(r)$ model with order parameter $r = 10$: The parameter $r$ is the maximum lag of the autoregressive process. The AR model generally predicts future behavior based on past data. The importance of the second as well as fourth coefficients show that there is some correlation between successive values in the time series for most of the users.

    \item The number of peaks in the time series with a support $(s)$ of at least $1$: A peak of support $s$ is defined as a sub-sequence in the time series where a value occurs that is greater than its $s$ neighbors to the left and to the right. When $s$ is set to $1$, this feature computes the number of peaks in the time series where a value is greater than its immediate neighbors. This feature can provide insights into the presence or intensity of localised fluctuations in the flow.
    
    \item The number of different continuous wavelet transform (CWT) peaks present in the signal for smoothing width of $1$: This feature was extracted from the time series by applying CWT using Ricker wavelet with width, $w=1$. This method simultaneously evaluates the signal in the temporal and frequency domains. The number of distinct peaks identified across the width scales considered can be quantified using these features. It can be used to compare the signals based on their peak characteristics.

    \item The value of partial autocorrelation function at a lag of $3$: The partial autocorrelation is a statistical measure that quantifies the linear relationship between a time series variable and its lagged values. In the context of our exhaled breath flow, the partial autocorrelation can provide insights into the temporal dependence and correlation structure of the breath velocity. This means that this feature can be useful in understanding the persistence or memory of the signal. It suggests that a strong linear relationship between the current flow state and its state $3$ time steps ago have been important for the classification of human subjects.

    \item Width of the multifractal spectrum ($\omega$) of the exhaled breath time series: $\omega$ describes the richness of the multifractality present in the time series, i.e., wider the range of singularity strength, richer the structure of the signal. The spectral width can implicitly represent the intensity or the level of turbulence present in the flow of exhaled breath. Turbulence is characterized by fluctuations in velocity at different scales. A wider range of turbulence scales is reflected by a wider spectral width, indicating a more turbulent flow. This might be attributed to factors such as extrathoracic constriction, or increased turbulence due to specific breath patterns or breath dynamics.

    \item Fourth coefficient of the autoregressive $AR(r)$ model with order parameter $r = 10$.

    \item The number of different continuous wavelet transform (CWT) peaks present in the signal for smoothing width of $5$.
    
    \item Kurtosis of the velocity time series calculated with the adjusted Fisher-Pearson standardized moment coefficient, $g2$: We know that Kurtosis is a higher-order statistical attribute of velocity signals. The heaviness of the tails of the probability density functions of normalized time series could be distinct for each user. This feature will help us in assessing the degree of deviation from the Gaussian distribution and provides evidence of skewed behaviour of the time series.
    
\end{enumerate}

\subsection*{Computational complexity of the algorithm}
\begin{figure}[!h]
\includegraphics[scale=0.46]{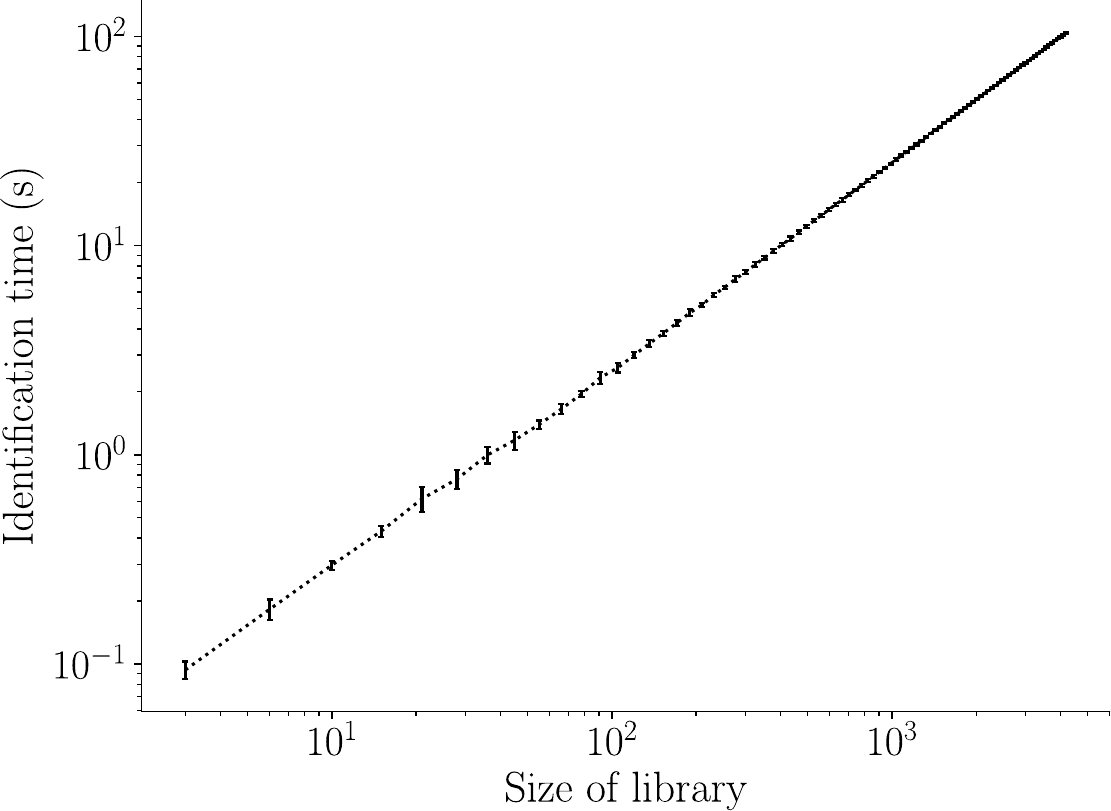}
\caption{{\bf Dependence of user identification time on the size of model library.}
Plot showing the linear relation of user identification time with the growth of model library. This is applicable to the ML based algorithms which include building of binary classifier models (also known as \textit{enrollment} in the context of biometrics). The error bars show $95\%$ confidence interval at every data point.}
\label{fig:identification_time_vs_library_size}
\end{figure}

Run-time of an algorithm is an extremely important factor for a real-time biometric system. It was generally observed that the size of the input feature set affects the amount of computational resources required to run an algorithm. It was observed that the hypothesis test based algorithm performs predictions faster than the machine learning based algorithm which is because the former is an instance-based classifier. Since the user identification algorithm depends on the number of users and in turn the number of models in the model library, the identification time per user was expected to scale up with the size of the library. The identification time was observed to show a linear relationship with the size of the library (of the form $y=ax$, with slope $a\approx 1$) as seen in the Fig \ref{fig:identification_time_vs_library_size}. The error bars show $95\%$ confidence interval at every data point.
            
One of the advantages of building an algorithm which uses $^{n}C_2$ binary classifiers instead of a single multi-class classifier is that it is massively parallelisable. As long as we have sufficient number of cores to run model loading and prediction, the parallelisation is possible. This significantly improves the computational time by several orders.

\section*{Conclusion}
We have provided evidence for the feasibility of a novel biometric system that works based on the turbulence information present in human exhaled breath. The use of a hot-wire anemometer for data acquisition allowed us to build a compact working setup. The faster response time of a constant temperature hot wire anemometer and the real-time computation in combination will possibly make the setup implementable as a biometric authentication system. Since the input of the exhaled breath-based biometric system is correlated with the internal morphology of the human body, it is impossible for a hacker to spoof-authenticate a user. This is because it is difficult to reconstruct an original time series and subsequently the binary classifier models that consolidate all the relevant features (biometric traits) of the true user. Preliminary studies carried out and presented in this work based on time series data from $94$ human subjects have shown promising results. We recommend the machine learning approach discussed in this work as a procedure to build a working user confirmation system as it produces good accuracy in confirming users. It achieved a true confirmation rate of nearly $100\%$, which is because of the ability of random forest models to capture complex decision boundaries between the classes. Although the dataset performs really well for a user confirmation algorithm, the real test of a biometric system comes in for the user identification algorithm, where the test user's identity is not revealed a priori. Building such an algorithm comes with more challenges and would require samples from a larger population to be evaluated. We recommend a multi-model approach for the user identification system, as discussed in this manuscript. The results from our study show that a user identification algorithm performs reasonably well with maximum precision and accuracy of $\approx 40\%$ each for optimum parameter settings. $39.4\%$ to $57.4\%$ of the users were correctly identified as at least the second highest voted users.

Our study reveals the possibility that a system built solely on the basis of the fluid dynamics of human exhaled breath could be a potential tool to understand the person-to-person variation in turbulent signatures of exhaled breath. This uniqueness in observed signature could potentially be correlated to the morphometric variation present in the extrathoracic airway. To make comments on the intricate structures within the upper respiratory tract, we might need experimental proof on cadaver models, or simultaneous imaging of upper tract along with the HWA data. Such a study would give us insights on how the structures exhibit considerable morphological diversity among individuals. While our study does not involve direct experimentation with throat morphology, it prompts consideration of how these morphological variations could contribute to the surprisingly unique turbulent signatures found in exhaled breath. Further investigation would give us better understanding on the relationship between these morphological traits and the distinct fluid dynamic signatures. For example, it is possible that the turbulence information can be correlated to occlusion in the extrathoracic passage and its nature, which is a major source of deposition of aerosolised therapeutics. Such an understanding will help us delve deeper into the area of personalised medicines.

\section*{Ethics statement}
The experimental data collection was carried out during January 2019. All experiments involving human subjects in this study were approved by the Institutional Ethics Committee (IEC) of the Indian Institute of Technology Madras (ref. IITM - IEC Protocol No. IEC/2018-03/MP/01). All volunteers who participated in this study had given their written consent. The data were analyzed anonymously.

\section*{Patent}
The authors declare there is a provisional patent on the discussed technology filed at Indian Patent Office, under the title `Exhaled breath based user authentication and diagnosis' (CONFIDENTIAL; - PATENT PENDING - 202241065024)

\section*{Acknowledgments}
The authors wish to acknowledge the financial support to Mr. Mukesh K, by the Ministry of Education, Government of India. The authors also acknowledge the HPCE, the Synchrony, and the SENAI of the Indian Institute of Technology Madras for providing the required high performance computing resources. The authors thank the NCCRD, IIT Madras, for providing the hot-wire anemometer setup and the calibration facility to carry out the experiments. The authors are also thankful to all the participants who volunteered to give their exhaled breath data for this study.



%
%
%

\end{document}